\documentclass[review,authoryear]{elsarticleLDB}
\usepackage[english]{babel}
\usepackage{natbib}
\usepackage{fancyhdr}
\usepackage[latin1]{inputenc}
\usepackage{amsbsy}
\usepackage{amsmath}
 \DeclareMathOperator*{\argmax}{arg\,max} 
 
\usepackage{amsfonts}
\usepackage{appendix}
\usepackage{rotating}
\usepackage{threeparttable}
\usepackage{longtable}
\usepackage{endnotes}
\usepackage{verbatim}
\usepackage{subfig}
\usepackage{float}
\usepackage{placeins}
\usepackage[a4paper,pdftex,colorlinks,citecolor=blue,linkcolor=blue,urlcolor=blue]{hyperref}
\usepackage{hyphenat}
\usepackage{breakcites}
\usepackage{verbatim}
\usepackage{booktabs}
\usepackage{fullpage}
\usepackage{setspace}
\usepackage[misc,geometry]{ifsym}
\usepackage{hyperref}
\usepackage[usenames, dvipsnames]{color}
\usepackage{chngcntr}
\usepackage{Sweave}
\usepackage{rotating}
\DefineVerbatimEnvironment{Scode}{Verbatim}{fontshape=\small}
\DefineVerbatimEnvironment{Statainput}{Verbatim}{
    fontsize=\tiny,
  fontfamily=courier,frame=single,framesep=2mm,
  label={Stata code},labelposition=topline} 
\DefineVerbatimEnvironment{Rinput}{Verbatim}{
    fontsize=\tiny,
  fontfamily=courier,frame=single,framesep=2mm,
  label={R code},labelposition=topline} 
\DefineVerbatimEnvironment{Pjkinput}{Verbatim}{
    fontsize=\tiny,
  fontfamily=courier,frame=single,framesep=2mm,
  label={Pajek code},labelposition=topline} 

\begin{document}

\title{Segregation with Social Linkages:
\\ Evaluating Schelling's Model with Networked Individuals}

\author[rc]{Roy Cerqueti}
\ead{roy.cerqueti@unimc.it}
\author[ldb]{Luca De Benedictis\corref{cor1}}
\ead{luca.debenedictis@unimc.it}
\author[vls]{Valerio Leone Sciabolazza}
\ead{valerio.leonesciabolazza@uniparthenope.it}

\cortext[cor1]{Corresponding author}

\address[rc]{DED, University of Macerata, Macerata, Italy; School of Business, London South Bank University, UK}
\address[ldb]{DED, University of Macerata, Macerata; Rossi-Doria Center, University Roma Tre; and Luiss, Italy}
\address[vls]{University of Naples Parthenope, Italy}

\begin{abstract}

\date{\today}
\small {This paper generalizes the original \cite{Sch1969, Sch1971a, Sch1971b, Sch2006}} model of racial and residential segregation to a context of variable externalities due to social linkages. In a setting in which individuals' utility function is a convex combination of a heuristic function \textit{� la} Schelling, of the distance to friends, and of the cost of moving, the prediction of the original model gets attenuated: the segregation equilibria are not the unique solutions. While the cost of distance has a monotonic pro-status-quo effect, equivalent to that of models of migration and gravity models, if \emph{friends} and \emph{neighbours} are formed following independent processes the location of friends in space generates an externality that reinforces the initial configuration if the distance to friends is minimal, and if the degree of each agent is high. The effect on segregation equilibria crucially depends on the role played by network externalities.

\begin{keyword}
Schelling's Segregation Model \sep Networks \sep Network externality.

\vspace{.4cm}
\small {\noindent \emph{Acknowledgements} We thank Tom Snyders and Luca Gori for valuable suggestions, and seminar and conference participants at ARS in Naples, and SIE-RSA in Arcavacata (CS). 
Roy Cerqueti and Luca De Benedictis acknowledge funding from the Italian Ministry of University and Research.}

\end{keyword}
\vspace{0.5cm}

\end{abstract}

\maketitle

\begin{center}{\large 30 December 2019}\end{center}

\newpage
\section{Introduction}\label{sec:intro}

The model of racial and residential segregation by Thomas Schelling is a prominent contribution to economic theory. 
Outlined in a series of papers \citep{Sch1969, Sch1971a, Sch1971b} and further elaborated in \cite{Sch2006}, the logic of Schelling's argument offers a simple and subtle intuition to racial residential segregation.\footnote{\ Quoting Schelling: ``My ultimate concern of course is segregation by color in the United States'' \citep[p.488]{Sch1969}.}
Schelling's main insight is that homophily - i.e. the tendency of people to associate with others who are similar to themselves -- in spite of individual preferences being characterized by the desire for integration, generates a completely unexpected social equilibrium where individuals tend to live within racially homogenous communities. Schelling's simple mechanism was originally framed as a heuristic process of strategic reasoning, in which people leave their residence whenever they are a too small minority in their neighborhood.\footnote{\ The same reasoning has been proposed using different setups and the main findings of the model were found consistent regardless of the definition of the neighborhood and its topography \citep{Tay1984}. For an account of the extensions of Schelling model to different contexts see \cite{Ioa2013} and \cite{EasKle2010}.}
Both versions of the Schelling model, the ``chequerboard model" - in which a fixed number of agents belonging to two different groups (the ``reds'' and the ``blue'' ones) choose where to locate on a grid (or on a line) as a consequence of their heuristic -- and the ``neighborhood tipping model'' - in which a small change in group composition of a neighborhood can put the heuristic in motion, and leads to an accelerating and irreversible dynamic process bringing to racial segregation -- are based on the same stringent logic in which observed \emph{macrobehavior} clashes with \emph{micromotives}: radical racial segregation is a robust and stable equilibrium even with individual homophily preferences over moderate racial integration.


The basic intuition of Schelling is supported by a large array of evidence on the persistence and ubiquity of racial and residential segregation across several countries: namely the US \citep{Cla1991, MobRos2001, Bat2013, Hen2014}, Europe \citep{vanTamdeVZwi2016}, and Israel \citep{HatBen2012}. However, Schelling's prediction of extreme racial spatial separation and ghettos rarely proved to be correct \citep{Eas2009}. On the contrary, the conformation of cities and urban areas, as far as segregation is concerned, is as various as it possibly could be. These aspects of reality, still unexplained by Schelling's work and its extensions, cast doubts on the ineluctability of what the model posits, and call for additional research.

Complementing Schelling's rational, this paper proposes a framework to unravel what forces might contrast dramatic population moves and produce the emergence of limited segregation equilibria. To this purpose, Schelling's heuristic is put in interplay with a positive cost of moving and, more importantly, the presence of social linkages \citep{JacRogZen2017}. Put differently, while Schelling considers the case when people identify only with one of their characteristics (e.g., ethnicity), we consider the possibility that the force of attraction determined by homophily is affected by different identities (e.g., ethnicity and social connections) and mitigated by the cost of moving. In what follows, we will show how this generalization provides a valuable insight on the formation of racially mixed neighborhoods as they result from both local inertia, generated by the incidence of moving costs, and network externalities, arising from the desire to live close to friends.

Of course, this is not the first case in which Schelling's model has been decontextualized, put in close analogy to concepts originated in other disciplines \citep{VinKir2006}, and extended so that its heuristic and other choice variables \citep{HenRosGre2005} could be included in a compact multivariate utility function. Among the many, \cite{SetSom2004} defines a framework in which agents care about both the racial composition and the affluence of neighborhoods, showing that, in a context of multiple equilibria, a reduction in inequality can be observed both with a \emph{rise} in segregation or a \emph{fall} in the level of affluence, depending on the speed of income convergence. \cite{Zha2004} instead, considers the existence of a housing market and agents chose to move according their heuristic and the price of an empty spot.\footnote{ \ Our model can naturally encompass extensions that expand the choice set of individuals, such as those of the above contributions.} 

In our setup, the interplay between moving costs and social linkages generates a network externality that has an effect on the segregation patterns. When the agent is able to move and simultaneously reduce the average distance from her friends, incentives to relocate to a segregated area will decrease for everyone. After the move, incentives will decrease for friends close to the agent, and will increase for all the others.

As a final note, we stress that 
even if the two Schelling's formalizations - the ``neighborhood tipping model'' and the ``chequerboard model''  - can be combined into one single theoretical framework (see, for instance, \cite{You2001} and \cite{Zha2011} and the excellent textbook treatment of the models by \cite{Ioa2013} and \cite{EasKle2010}), this approach would add unnecessary complexities to our work. By contrast, we aim at keeping the details of this work as simple as possible and, as in \citep{Pan2007}, we build our contribution around the original Schelling ``chequerboard model,'' in which the number of individuals and their characteristics are well defined and the number of neighborhoods is higher than one. 

Albeit analytically  different, this paper has strong connections with the literature originated by the seminal contribution of George Akerlof and Rachel Kranton on the Economics of Identity \citep{AkeKra2000} and summarized and popularized in \cite{AkeKra2012}. In their framework, agents make their choice according to the standard utility maximization apparatus - market prices, individual income and preferences - \emph{and} an identity variable, depending on individual characteristics (e.g. being female, being white, being homosexual) and self-image, on the social context and on the behavior of others. Choices depend therefore on economic incentives and identity issues. In our context, the identity of an individual will depend on a single characteristic (e.g., color) and on a social norm determining the degree of homophily in Schelling's heuristic. The market equilibrium in residential location and the subsequent level of social segregation will depend on economic factors (the cost of moving), identity (color and social norms) and social linkages (friends, their choices and the willingness to minimize the spatial distance between individual residence and the residence of friends). 

The remainder of the paper is organized as follows. In Section 2, we show how Schelling's model can be expanded so to include the effects played by network externalities and moving costs. The components of this model are further explicated in Section 3, where we detail how they can be used to investigate segregation equilibria in a simulated setting. Section 4 presents the results of such simulations. Finally, Section 5 concludes. Before moving to Section 2 however, we briefly recall Schelling's reasoning {through an illustrative example} for the reader's convenience, and discuss the rational underlying it.

\subsection{An example: Schelling ``chequerboard model" with and without social linkages}\label{sec:ex}

\begin{figure}[H]
  \caption{The Schelling ``chequerboard'' model.}
  \centering \vspace{-1.5cm}
    \centerline{\includegraphics[width=0.7\textwidth]{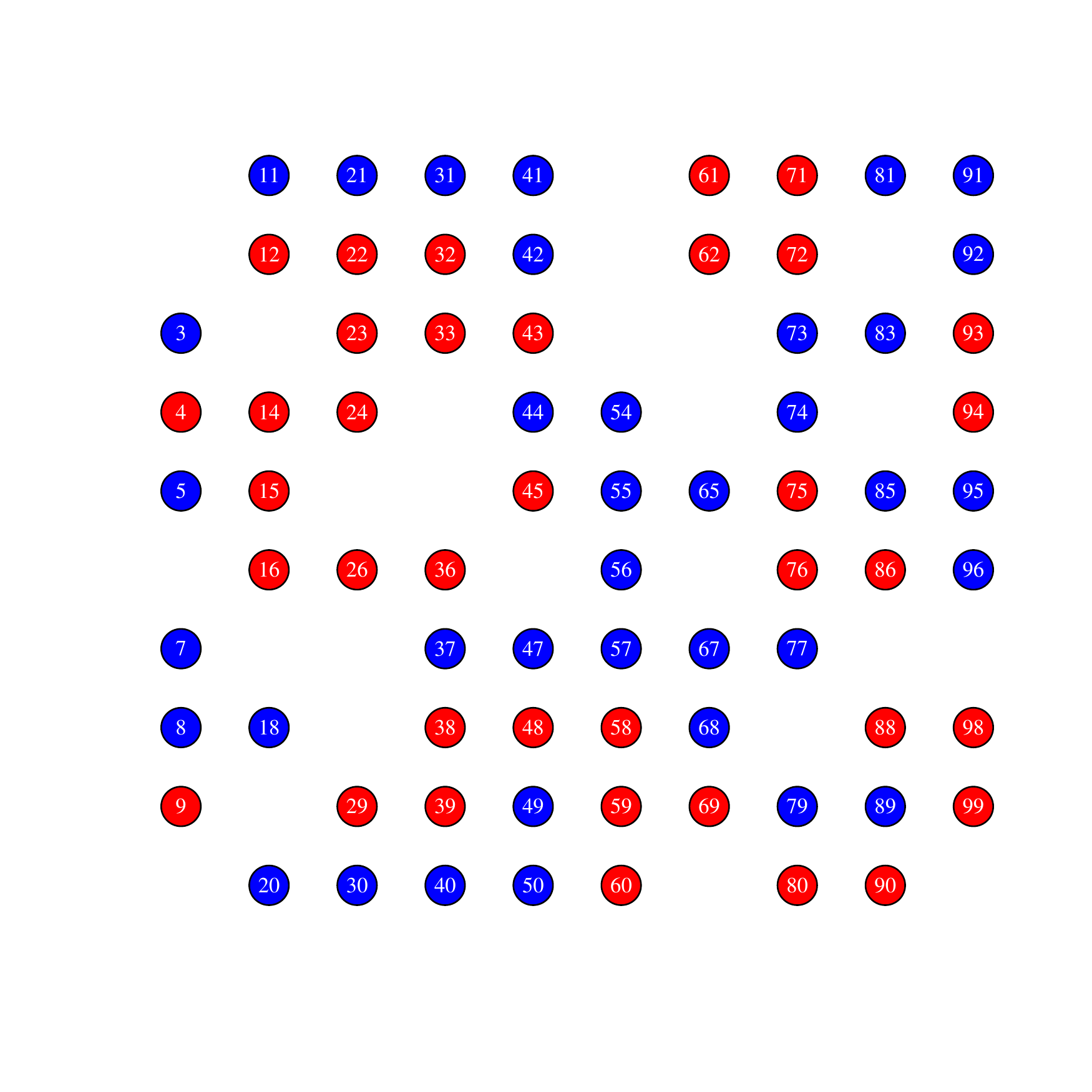}}
  \label{Fig:Schelling1}
\end{figure}

Consider Figure \ref{Fig:Schelling1}, in which 74 individuals are located on a 10$\times$10 grid. 37 of them are ``reds", and 37 are ``blues." 
The 26 spots that are not occupied neither by reds nor by blues remain empty and can be occupied in a subsequent round of play. Finally, the initial position of each individual is random.

Each individual has up to eight adjacent neighbors and has a preference over homophily. Following a heuristic process, the generic individual remains in her location as long as at least {a share $x$} of her neighbors are of the same type as she is. In this example, we set $x=1/3$. Let's now consider individual 23 in Figure \ref{Fig:Schelling1}.\footnote{ \ In Schelling's original settings all agents unsatisfied of their location simultaneously put their name on a list, and then they move according to an empty space following some arbitrary order. Successively, a new list is drawn and the process repeats until no agents have further incentive to move. Here, this process is simplified to avoid unnecessary complexities, and we select one agent randomly.} 23 is red and is surrounded by six reds and two empty spots. Since empty cells are excluded from the heuristic accounting, individual 23 compares 6/6 > 1/3 and she decides not move from her position on the grid. Individual 45, instead, is red as well but is surrounded by four blues, one red and three empty spots. Since 1/5 < 1/3 she will move from her position on the grid. Similarly, all unhappy individuals will 

\vspace{-1.5cm}
\begin{figure}[H]
  \caption{The Schelling ``chequerboard'' model with networked individuals.}
  \centering \vspace{-1.5cm}
  \centerline{\includegraphics[width=0.7\textwidth]{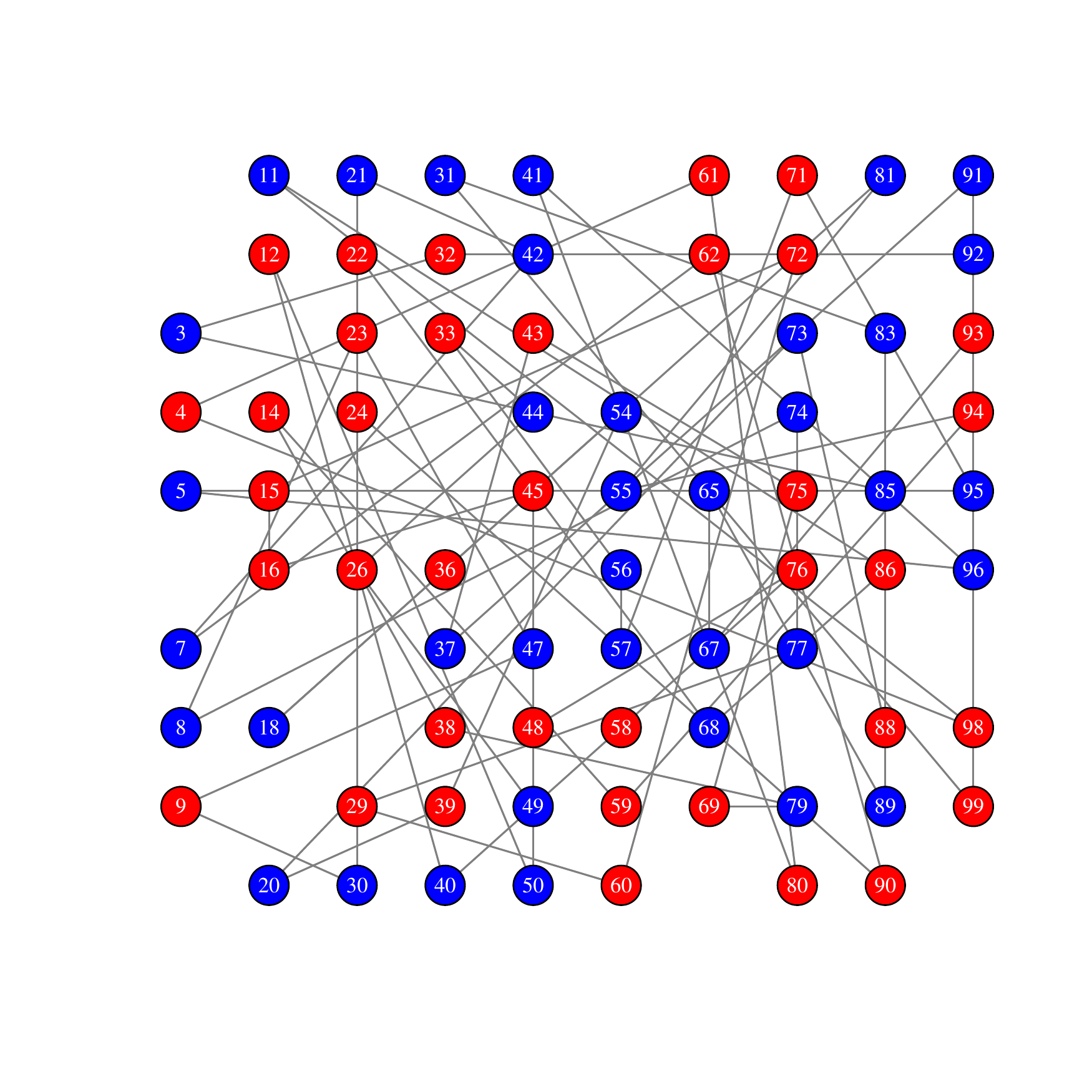}}
    \label{Fig:Schelling2}
  {\tiny \textbf{Note}: The setup is similar to the one in \cite{Sch1969, Sch2006}.}
\end{figure}
move to an empty spot that satisfies the ``at least one-third'' heuristic. Segregation will rapidly emerge.\footnote{ \ All possible multiple-equilibria are stable and segregation can be reduced only introducing a distaste for uniformity, corresponding to a heuristic > 1/3 \emph{and} < 7/8, as an example. The formulation is analogous to \cite{Atk1970} inequality aversion.}

In Figure \ref{Fig:Schelling2}, individuals are located in the same places as in Figure \ref{Fig:Schelling1}. They are also similarly split in two different groups, the ``reds'' and the ``blues," but, with respect to Schelling's original formulation, they are also characterized by a fixed number of time-invariant friends. In this setting, individual 23 is friend with 47 and 8, and individual 45 is friend with 54 and 36. Conditional on the heuristic, they both prefer to live close to their friends, who thus modifies their individual choices. Individual 23 will tend to move to get closer to 8 and 47. On the contrary, individual 45 will now tend to stay, since her friends live nearby. The effect of friends is therefore non-monotonic as far the choice of moving is concerned, and generates a variable network externality on friends: e.g. individuals 47 and 8 will tend to stay where they are after the move by 23.

The main elements of our reasoning will be detailed, generalized and formalized in the subsequent model.

\section{Schelling model with networked individuals}\label{sec:model}
For the sake of clarity, we proceed in a stepwise form.

\subsection{The framework}\label{subsec:framework}

Consider a squared grid with side $n$, hence containing $n^2$ cells.
We denote the set of the cells as $V$.

Each cell of the grid is identified by its row and column, so that
$v_{rc} \in V $ is the cell at the intersection of row $r$ and
column $c$, for each $r,c=1,\dots,n$. For the sake of simplicity and
when needed, we will scall $v_{rc}=v$, where $v=(r-1)n+c$. Such a
definition of $v$ labels with a real positive number all the elements of $V$, and
it results that $v=1, \dots, n^2$. Hereafter, without loss of generality,
we will use equivalently $v \in V$ and $v\in \{1, \dots, n^2\}$.

We assume that the grid is suitably recombined at its boundary to be
shaped as a torus, so that - for every $0 < h < n$ - each $v_{(n+h)c}=v_{hc}$ and
$v_{r(n+h)}=v_{rh}$, for each $r,c,h=1, \dots, n$.

We introduce two different populations in the grid: the red individuals
and the blue ones. For every individual being blue or red is an immutable trait. The set collecting the reds and the blues are
denoted by $\mathcal{A}$ and $\mathcal{B}$, with $|\mathcal{A}|=A$
and $|\mathcal{B}|=B$, respectively. We assume that the entire
population belongs to the set $\mathcal{P}=\mathcal{A} \cup
\mathcal{B}$. Each cell of the grid can be occupied by an element of
$\mathcal{P}$ or it can be empty. Therefore, we have $A+B<n^2$. The
generic elements of $\mathcal{A}$ and $\mathcal{B}$ will be denoted
by $a$ and $b$, respectively. We will refer to them as
\textit{agents}, and denote the generic agents as $i,j \in
\mathcal{P}$.

All agents, also those sharing the same color, differ for the
social connections they have. The social connections of agent $i$
are represented by a star-shaped graph with $i$ playing the role of
the central hub. By merging together all the social connections of
the agents in the grid, we have a unified graph composed by the
individual $A+B$ (possibly interconnected) star-shaped subgraphs. We
denote by $\mathbf{G}=(\mathcal{P},E)$ the graph in the grid, where
$E$ represents the set of the edges of $\mathbf{G}$.

Specifically, we say that $(i,j) \in E$ if and only if there is a
social connection between agent $i$ and agent $j$, for each $i,j \in
\mathcal{P}$. We define
$$
\mathcal{P}_i:=\{j \in \mathcal{P}|(i,j) \in E\}, \qquad \forall\,i
\in \mathcal{P}.
$$
The set $\mathcal{P}_i$ is the \textit{friendship set} of the agent
$i$, and $j \in \mathcal{P}_i$ is said to be \textit{friend} of $i$.
We assume that friendship disregards colors and thus, in general,
$\mathcal{P}_i \cap \mathcal{A} \not= \emptyset$ and $\mathcal{P}_i
\cap \mathcal{B} \not= \emptyset$, for each agent $i \in
\mathcal{P}$.

Moreover, links are assumed to be undirected, so that $ j \in
\mathcal{P}_i$ if and only if $i \in \mathcal{P}_j$, for each $i, j
\in \mathcal{P}$. At the same time, links are not weighted. Briefly, $i$ and $j$ are \textit{friends}.


\subsection{Agents' preferences in the grid}

Agents move in the grid, and their movement is driven by their level of satisfaction in occupying a given cell of the grid. Here, we discuss what are the criteria driving the preferences of an agent for a cell, while in the next subsections we will present agents' movement rule and the random assignment mechanism used to configure their initial position in the grid.

\begin{itemize}
\item \textit{Color.} According to the classical Schelling model, we
assume that agents prefer to be surrounded by people of the same
color, and have a disutility in being surrounded by a majority of
agents showing a color different from her own one. We formalize
this assumption.

Given $\bar{r},\bar{c} =1, \dots, n$ and the cell
$\bar{v}=v_{\bar{r}\bar{c}} \in V$, we define the \emph{neighborhood
set of} $\bar{v}$ by
\begin{equation}
\label{Ni} N_{\bar{v}}:=\{v_{rc} \in V\setminus \{\bar{v}\}|(r,c)
\in \{\bar{r}-1,\bar{r},\bar{r}+1\} \times
\{\bar{c}-1,\bar{c},\bar{c}+1\}\}.
\end{equation}
By construction, we have that $|N_{\bar{v}}|=8$. For each $\bar{v}
\in V$, we introduce two variables $\xi^A_{\bar{v}}$ and
$\xi^B_{\bar{v}}$ which describe the colors of $N_{\bar{v}}$ as
follows:
$$
\xi_{\bar{v}}^A:=\frac{|N_{\bar{v}} \cap \mathcal{A}|}{8},\qquad
\xi_{\bar{v}}^B:=\frac{|N_{\bar{v}} \cap \mathcal{B}|}{8}.
$$

In general, $\xi_{\bar{v}}^A+\xi_{\bar{v}}^B \leq 1$.

For any agent $i \in \mathcal{P}$, the color-based normalized
utility function of $i$ in being located in the cells in $V$ is
$U^{color}_{i}:V \rightarrow [0,1]$ such that
$U^{color}_{i}(\bar{v})$ is not decreasing (not increasing,
respectively) with respect to $\xi_{\bar{v}}^A$ when $i \in
\mathcal{A}$ ($i \in \mathcal{B}$, respectively).

\item \textit{Friendship.} Agents prefer to locate close to their
friends. To explain this extremely reasonable assumption, the
introduction of a distance measure between couples of cells
belonging to the squared grid is required. Without losing
generality, we employ the normalized Manhattan distance $d:V^2
\rightarrow [0,1]$ defined as

\begin{equation}
\label{dij} d(v_{r_1,c_1},v_{r_2,c_2})=\left\{
\begin{array}{ll}
\frac{|r_1-r_2|+|c_1-c_2|}{n}, & \hbox{if $n$ is pair;} \\
\frac{|r_1-r_2|+|c_1-c_2|}{n-1}, & \hbox{if $n$ is odd.}
\end{array}
\right. ,
\end{equation}
where $r_1,r_2,c_1,c_2=1, \dots, n$ and $v_{r_1,c_1},v_{r_2,c_2} \in
V$ and $n$ in the normalizing factor.

With a reasonable abuse of notation, we will refer to $d(i,j)$ as
the distance between two agents $i,j \in \mathcal{P}$, to be
intended as the distance between the cells occupied by $i$ and $j$
and in accord to formula (\ref{dij}). In the same way, we will refer
to $d(i,v)$ as the distance between the agent $i \in \mathcal{P}$
and the cell $v \in V$, which represents the distance between the
cell occupied by $i$ and the cell $v$. If agent $i$ occupies the
cell $v$, then $d(i,v)=0$.

Notice that when $i$ and $j$ are friends, then $d(i,j)$ is the
Manhattan distance between the cells representing the extremes of
the edge $(i,j)$. Moreover, by construction, $d(i,j) \in [0,1]$.

Given an agent $i \in \mathcal{P}$, we can provide a measure of the
friendship set $\mathcal{P}_i$. Clearly, such a measure depends on
the specific cell $v \in V$ from which $i$'s friends are observed.
We denote such a measure as $\Delta(v,\mathcal{P}_i)$.

The measure $\Delta$ describes how an agent changes her
closeness to her friends by modifying her position in the
graph. As already mentioned above, we naturally assume that agent
$i$, located in cell $v$, feels to be more satisfied about her
connections (and surrounded by friendship) as the value of the
$\Delta(v,\mathcal{P}_i)$'s is lower.\footnote{ \ In the simulation in Section \ref{sec:simulation}, we will provide a specific form of the $\Delta$ metric.}








For any agent $i \in \mathcal{P}$, the friendship-based normalized
utility function of $i$ in being located in the cells in $V$ is
$U^{friend}_{i}:V \rightarrow [0,1]$ such that $U^{friend}_{i}(v)$
is not increasing with respect to $\Delta(v,\mathcal{P}_i)$, for
each $v \in V$.

\item \textit{Cost of moving.} Agents spend an effort in moving from a
cell to another one, and such effort generates a disutility which
does not decrease as the distance from the involved cells increases.

Thus, for any agent $i \in \mathcal{P}$, the distance-based
normalized utility function of $i$ in moving from her position to
the cells in $V$ is $U^{moving}_{i}:V \rightarrow [0,1]$ such that
$U^{moving}_{i}(v)$ is not increasing with respect to $d(i,v)$, for
each $v \in V$.
\end{itemize}

The utility function $U_i:V \rightarrow [0,1]$ -- which drives the
movement of the agents in the grid -- is a convex combination of the
utility functions described above, so that
\begin{equation}
\label{Ui} U_i(v):= \alpha_c U^{color}_{i}(v)+\alpha_f
U^{friend}_{i}(v)+ \alpha_d U^{moving}_{i}(v),
\end{equation}
where $\alpha_c, \alpha_f, \alpha_d \geq 0$ such that
$\alpha_c+\alpha_f+\alpha_d=1$. The selection of the values of the
$\alpha$'s explains in a clear way the relevance of color,
friendship and movement effort in implementing the decision of
moving from a cell to another one.

We stress that utilities have extreme zero (resp. unitary) value at the lowest possible
(resp. highest possible) level of satisfaction. The range of the utility functions
plays a key role here. In fact, the values of the utility function
over the cells can be compared, and such comparison will drive the
allocation procedure of the agents in the cells of the grid.


\subsection{The movement rule}

The preferences described above represent the drivers of the
movement of the agents in the grid. In this section, we describe how
agents change their positions among the cells.

Firstly, and without losing of generality, we assume that movements
are on a sequential basis, that is the location process of
the agents is with discrete time: i.e. time $t \in
\mathbb{N}$. The way in which agents move is straightforward: any moving
agent changes her location cell with an empty one. In so doing,
the graph $\mathbf{G}$ does not change, in the sense that
friendships are constant over time. The component of the model which
changes is represented by the set of the cells of the grid occupied
by an agent and the empty ones. Given a time $t \in \mathbb{N}$, we
denote the set collecting the empty cells at time $t$ by $Em_t$ and
the set of the occupied ones by $NEm_t$, so that $V=Em_t \cup
NEm_t$, for each $t \in \mathbb{N}$. By construction, $Em_t \cap
NEm_t = \emptyset$.

We assume that only one agent moves each time.\footnote{ \ The extension
to the case of simultaneous movements of agents can be, of course,
presented. However, it does not add much to the single-moves case.}

For a fixed time $t \in \mathbb{N}$, we denote the
\textit{configuration of the grid at time $t$ by}
$V_t=(EM_t,NEm_t)$. The agents start to move at time 1. Thus, $V_t$
represents the situation of the grid at the $t$-th move -- i.e.:
after the moves of $t$ agents -- and identifies empty and occupied cells at time $t$.

We denote by \textit{initial configuration} the grid at time $t=0$,
i.e. $V_0$. 

Of course, over time, the value of the utility functions
of each agent in the cells of the grid may change. In fact,
the passage from a configuration to another one is obtained by
changing the position of one agent, and this might lead to a
variation of the constitutive terms of the utility function $U_i$ in
(\ref{Ui}) for some $i \in \mathcal{P}$. For any $i \in \mathcal{P}$ and $t \in \mathbb{N}$, we denote the
utility function of agent $i$ of being located in the cell $v$ at
time $t$ as $U_i(v,t)$.

We are now in the position of describing the mechanism of the
movement.

An agent moves only if the change of her position leads to an
improvement of her utility. The first moving agent is the
saddest one, i.e. the one with the lowest level of utility of being
located in her current position. The moving agent selects the
empty cell which provides her/him the maximum level of utility.\footnote{ \ A movement rule where, at each stage, one agent is randomly selected and asked to choose a best-response was previously used by \cite{Pan2007}.}

If more than one empty cell maximizes the utility, then the occupied
one is randomly selected among the utility maximizing cells.
Analogously, if more than one agent is in the position of moving,
then the moving one is selected according to a uniform distribution
over the available alternatives.

As an intuitive premise, we assume that any agent $i \in
\mathcal{P}$ has a complete information at time $t$ of the utility
function $U_i(v,t)$, where $v$ is the cell occupied by $i$ at time
$t$ or $v \in Em_t$. Moreover, each agent is also aware about the
peculiar distribution at time $t$ of the agents belonging to
$\mathcal{A}$ and $\mathcal{B}$. The agent moving at time $t$ will
be labeled as $j^\star_t$, whose definition is obtained according to
the following two conditions:
\begin{equation}
\label{jstar} \left\{
  \begin{array}{l}
   j^\star_t \in {\rm argmin}\left\{U_j(\bar{v},t) \,|\,j \in \mathcal{P}\text{ located in }\bar{v} \in NEm_t\right\}; \\
\exists \, v \in Em_t \,|\,U_{j^\star_t}(v,t) \geq
U_{j^\star_t}(\bar{v},t)
  \end{array}
\right.
\end{equation}
The two conditions in (\ref{jstar}) must be jointly verified: the
former one indicates that the moving agent is that with the lowest level of utility exerted from her position; the latter one specifies that the agent changes her position only if she will improve her utility.

The mechanism admits hypothetically simultaneous movements of
different agents at time $t$, meaning that system (\ref{jstar}) can be
satisfied by more than one agent. By assuming that $Q$ agents
satisfy (\ref{jstar}), with $Q=1,\dots, A+B$, we will label the $Q$
moving agents by $j^\star_{1,t}, \dots, j^\star_{q,t},\dots,
j^\star_{Q,t}$. A random extraction of one element of the set $\{1,\dots,Q\}$
identifies univocally $\bar{q} \in \{1,\dots,Q\} $ such that the
moving agent is $j^\star_t=j^\star_{\bar{q},t}$.

Once condition (\ref{jstar}) is satisfied and the moving agent
identified, we implement the selection of the empty cell where agent
$j^\star_t$ moves. As already mentioned above, this selection is driven by the agent's utility function. We model the utility of an agent to move to an empty cell $v^\star_{t}$ as follows:
\begin{equation}
\label{estar}
   v^\star_{t} \in \Phi_{t}:={\rm argmax}\left\{
U_{j^\star_t}(v,t) \,|\,v \in Em_t \text{ and }
U_{j^\star_t}(v_{t-1},t) < U_{j^\star_t}(v,t)\right\},
\end{equation}
with the conventional agreement that $v^\star_{t}=v_{t-1}$ if
$\Phi_{t}= \emptyset$.

If $|\Phi_{t}|=C>1$, then there are $C$ different cells in $Em_t$
which are indifferent -- in terms of the utility function -- for the
moving agent $j^\star_t$. We will label such $C$ utility maximizer
cells by $v^\star_{1,t}, \dots, v^\star_{c,t},\dots, v^\star_{C,t}$.
In this case, we extract randomly one element of the set
$\{1,\dots,C\}$, and identify accordingly $\bar{c} \in \{1,\dots,C\}
$ such that the selected empty cell is
$v^\star_t=v^\star_{\bar{c},t}$.

The movement of $j^\star_t$ induces a new configuration of the grid,
which passes from $V_{t-1}$ to $V_{t}$. The movement process then
continues according to this mechanism. A new moving agent
$j^\star_{t+1}$ is identified, which changes her original
position with the empty cell $v^\star_{t+1}$.

The process stops at the first time in which all the agents cannot
improve their utility when moving from their position to an empty
cell. Clearly, the process can also never stop. For this reason, in
the numerical experiments we will stop the process also
in presence of loops, i.e. in the circumstance of an agent moving
iteratively back and forth between two cells.

\section{Simulation}\label{sec:simulation}

The dimensionality of the problem requires the model to be solved by simulation \citep{Axe1997, VelZam2015}. The basic elements of the model, described in Section \ref{sec:model}, are articulated through 
computation\footnote{ \ All computation is programmed in R and is available upon request for replication purposes.} and mimic a process that, given basic exogenous conditions (i.e. the spatial dimension, the density of agents, the relative share of the two populations) and the social context \citep{AkeKra2000}, begin with agents' moving choices and ends to the determination of stable equilibria, corresponding to a certain level of individual and social utility and of racial residential segregation. The simulation procedure can be seen as the numerical analogous to an experimental procedure \citep{BonSan1997}, where the parameters of the model are shocked to check the response of the system and evaluate the resulting levels of utility and segregation.

In this section we details the main aspects of our simulation setup. 

\subsection{Generating starting positions and network connections}
We consider an $n \times n$ torus with $n=10$, and impose that $A=B= 37$, and 26 cells are left empty. Agents' initial configurations are established at random. In order to avoid spurious
results, the initial configurations $V_0^{(h)}$ is permuted $H$ times by using just as many random seeds, so that the $h$-th seed generates the $h$-th initial configuration of the grid $V_0^{(h)}$, with $h=1, \dots,
H$. We set $H = 100$, to get a distribution of results.  The moments of said distribution are then summarized and plotted systematically.

Next, we create the graph $\mathbf{G}$. For consistency with the literature of social networks (\cite{New2010, Bar2016} and \cite{Goy09} and  \cite{Jac2010} for applications to economics), we denote by $k=0,1,\dots, 73$  the degree centrality of each node, which in this case it is assumed to be fixed, i.e. we assume a common degree $k$ for the hub of the star-shaped subgraphs. By considering the random seed of initial configuration, $h$ and the degree $k$, we define $\mathbf{G}:=\mathbf{G}_{h,k}=(\mathcal{P},E_{h,k})$ as
composition of the regular star-shaped graphs for the agents in
$\mathcal{P}$.  Clearly, the cardinality of the set of arcs $E_{h,k}$ depends on $k$, so that e.g. graph $\mathbf{G}_{h,2}(m)$ will
have $74\times 2$ arcs, and all the agents will have degree centrality equal to 2 -- i.e., each agent has two friends. For the sake of simplicity, and without loss of generality, we assume that $E_{h,k_1} \subset E_{h,k_2}$ with $k_1<k_2$: i.e. an arc in a graph $\mathbf{G}_{h,k_1}$ is also an arc in $\mathbf{G}_{h,k_2}$.
Consequently, $\mathbf{G}_{h,k_1}$ can be regarded as a subgraph of $\mathbf{G}_{h,k_2}$ when $k_1<k_2$.

\subsection{The utility function}
We now introduce the functional form of formula (\ref{Ui}) assumed by the simulations.

To this purpose, we begin with the component of agent's utility exerted from their location, $U^{color}$. We introduce a pre-specified threshold $x \in (0,1)$ which drives Schelling's heuristic. We label $x$ as \textit{Schelling's threshold}. Specifically, $x$ determines the maximum density of same-color neighbors required by $i$ to be satisfied by her residential location, so that
\begin{equation}
\label{ucol-ex}
U_i^{color}(\bar{v})=U_i^{color}(\bar{v};x)=\left\{
\begin{array}{ll}
\xi_{\bar{v}}^A \cdot 1_{\{\xi_{\bar{v}}^A
	> x\}}, & \hbox{if $i \in \mathcal{A}$;} \\
\xi_{\bar{v}}^B \cdot 1_{\{\xi_{\bar{v}}^B
	> x\}}, & \hbox{if $i \in \mathcal{B}$;}
\end{array}
\right.
\end{equation}
for each $\bar{v} \in V$ and $1_\bullet$ is the indicator function of the set $\bullet$. Formula (\ref{ucol-ex}) gives that when $x = 0$, $i$ is ``color-blind'' and does not benefit from any change of location when looking at the color of the surrounding agents. By contrast, when $x = 1$, $i$ will enjoy any marginal increase in the number of same-color neighbours.

Next, we define the utility obtained from friendship, $U^{friend}$, as:

\begin{equation}
\label{ufriend-ex}
U_i^{friend}(\bar{v}) = 1-\Delta(\bar{v},\mathcal{P}_i),
\end{equation}
where $\Delta(\bar{v},\mathcal{P}_i)$ is the average Manhattan distance in (\ref{dij}) between $i$ and her friends, i.e.:
\begin{equation}
\label{ufriend-Delta}
\Delta(\bar{v},\mathcal{P}_i)=\frac{1}{k}\sum_{j \in \mathcal{P}_i}d(i,\bar{v}).
\end{equation}

Then, we formalize the cost of moving, $U^{moving}$, as:
\begin{equation}
\label{umov-ex}
U_i^{moving}(\bar{v}) = 1 - [\gamma \bar{c} + (1- \gamma ) (1 - \bar{c})  (1 - d(i,\bar{v}))]
\end{equation}
Where $\bar{c}\in (0,1)$ is the cost of moving between two adjacent cells and $\gamma \in \{0,1\}$ determines whether costs are fixed ($\gamma = 1$) or variable ($\gamma = 0$).

Finally, we turn to the responsiveness of $i$'s utility to a change in $U_{color}$, $U_{friend}$ and $U_{moving}$, and we set:
\begin{equation}
\label{param-ex}
\begin{array}{l}
\alpha_{c} = \beta \alpha\\
\alpha_{f} = \beta(1-\alpha)\\
\alpha_{d} = 1 - \beta\\
\end{array}
\end{equation}
where $\alpha \in [0,1]$ and $\beta \in [0,1]$.

As a result, we can re-write formula \ref{Ui} by using (\ref{ucol-ex})-(\ref{param-ex}) and explicitating the Schelling's threshold $x$ of the utility dependent on the color as follows:
\begin{equation}
\label{UF}
U_{i}(\bar{v}) = U_{i}(\bar{v};x)=\beta\bigg(\alpha U_i^{color}(\bar{v};x) + (1-\alpha) U_i^{friend}(\bar{v})\bigg)+ (1-\beta) U_i^{moving}(\bar{v}). 
\end{equation}

\subsection{Measures of segregation}
Measuring the level of segregation adds an additional layer of complexity to our analysis. The level of agents' dissimilarity in a location is traditionally studied at the neighborhood level, where the boundaries of the residential area are exogenously defined \citep{MasDen1988}. By contrast, here we are interested in the agent's level of segregation, which is endogenously determined by the agent's own location. For this reason, we prefer over standard metrics of segregation \citep{MasDen1988}, the measurement provided by the Freeman Segregation Index ($FSI$) \citep{Fre1978}, and the Moran's $I$ index \citep{Mor1950}, which allow to assess the degree to which individuals with similar attributes are located close to one another when the neighborhood is self-referenced.

\cite{FagValVri2007} and \cite{Mel2017} show that $FSI$ can be profitably used to study agents' contiguity in a grid-like topological space, such as a torus, by using a graph perspective. Consistent with this approach and according to the model outlined above, we consider two groups of agents -- the red and the blue ones -- whose cardinalities are $A$ and $B$, respectively, and label all the agents with an integer $i=1, \dots, A + B$. We consider the friendship graph $\mathbf{G}$ and introduce its randomized version $\mathbf{G}_h$, where $h=1, \dots, H$ is the seed, as follows: agents represent the nodes and the generic entry $w^{(h)}_{i,j}$ of the symmetric adjacency matrix $W_h$ is 1 when the two agents $i$ and $j$ are adjacent in the grid -- according to formula (\ref{Ni}) -- and 0 otherwise. 
The maximum number of arcs $\bar{N}_{E}$ in $\mathbf{G}_h$ does not depend on $h$ and is given by:
\[
\bar{N}_{E} = \binom{A+B}{2} = \frac{(A+B)(A+B-1)}{2},
\]
while the maximum number of cross-group arcs is $AB$. 
Consequently, if agents are "color blind" and arcs are generated independently of each other, 
the probability of a given arc being a cross-group arc is:
\[
p = \frac{AB}{\bar{N}_{E}} = \frac{2AB}{(A+B)(A+B-1)}
\]
As a consequence, given the number of arcs in $\mathbf{G}$, say $N_{E}$, the expected number of cross-group arcs is $E(N^{*}_{E})$ defined as follows:
\[
E(N^{*}_{E}) = N_{E} p
\]
Given this formalization, $FSI$ is obtained as:
\begin{equation}
FSI = \argmax\left\{0, \frac{E(N^{*}_{E})- N^{c}_{E}}{E(N^{*}_{E})}\right\},
\label{eqn:FSI}
\end{equation}
where $N^{c}_{E}$ is the observed number of cross-group arcs. $FSI$ varies between 0 and 1, and it may be interpreted as the proportion by which the expected number of cross-group links is reduced in observation. Specifically, a value of 0 indicates full integration, while a value of 1 corresponds to full segregation. Put differently, when $FSI = 1$, there is a positive autocorrelation in the spatial distribution of agents: i.e. blue and red nodes are located in different and non-contiguous areas. On the contrary, when $FSI = 0$, there is no autocorrelation present, i.e. blue and red nodes are distributed at random in the torus.

We complement the FSI, with the Moran's $I$ index, which is calculated as:
\begin{equation}
I = \Big( \frac{A+B}{\sum_{i = 1}^{A+B} \sum_{j = 1}^{A+B} w_{i,j}} \Big)\Big(
\frac{\sum_{i = 1}^{A+B} \sum_{j = 1}^{A+B} w_{i,j} (z_{i} - \bar{z}) (z_{j} - \bar{z})}
{\sum_{i = 1}^{A+B} (z_{i} - \bar{z})^{2}}
\Big)
\end{equation}
where $z_{i}$ is equal to 1 if $i$ is red, and 0 if $i$ is blue, and $\bar{z} = \frac{\sum_{i = 1}^{A+B} z_{i}}{A+B}$. The values of the Moran's $I$ range in the interval $[-1,1]$. When $I=-1$, then one has the maximum possible negative spatial autocorrelation: e.g. red agents are adjacent only to blue agents. When $I=1$, there is the maximum level of positive spatial autocorrelation: e.g. each red agent is surrounded by red agents.

It is interesting to note that $FSI$ and Moran's $I$ are both correlated with another well-known index, that is \cite{Gea1954}'s metrics $G$, a measure of spatial autocorrelation which goes from 0 (maximum level of positive autocorrelation) to 2 (maximum level of negative autocorrelation). Specifically, $G$ is equal to the reciprocal of $FSI$, when $G \in (0, 1)$ \citep{Fre1978}, and it is inversely correlated with $I$ \citep{Ans1995}. By virtue of this, $FSI$ and Moran's $I$ are positively correlated when $I \geq 0$. However, they are not identical. The former is more sensitive to local patterns of segregation, while the latter is better suited for detecting segregation at the global level. The reason is that $FSI$ mimics the measurement of $G$, and \cite{Ans1995} has shown that $G$ is more susceptible than $I$ to local structures of segregation. We expect this difference to be preserved also when comparing $FSI$ and $I$.

\section{Simulation experiments, results and discussion}
\label{sec:results}
This section presents the simulation experiments, and discuss the results of our investigation. In order to disentangle the effect exerted by the different components of the utility function, as formalized in formula (\ref{UF}), a stepwise approach is adopted. First, we test plain Schelling's model, without moving costs and friendship connections. Next, we include the presence of moving costs, and then the presence of agents' connections. Finally, we present the case of Schelling's model and friendship, without moving costs, and in the extreme case of maximum Schelling's threshold $x=1$. 

The results of the simulation are assessed by looking at six different outcomes: i) the number of iterations required by the model in order to converge and produce the final configuration; ii) the number of agents who left their initial position during the iteration process; iii) the $FSI$ and iv) the Moran's $I$ index of the torus when model convergence is reached; v) the average and the vi) total social welfare of the agents at the end of the iteration process. Formally, v) and vi) are the average and the sum of individual agents' utilities in the final configuration of the torus, as obtained from formula (\ref{UF}). For each model setting, we report the average and the standard deviation of these metrics obtained $H$ different random initial configurations of the torus.

\subsection{The Schelling's model}

In this first array of experiments, moving costs and friendships are sorted out from the definition of the utility function in (\ref{UF}), i.e. $\beta=\alpha=1$. The analysis is focused on agents' responsiveness to the variation of the Schelling's threshold $x$.

\begin{figure}[H]
    \begin{center}
        \includegraphics[width=0.8\textwidth]{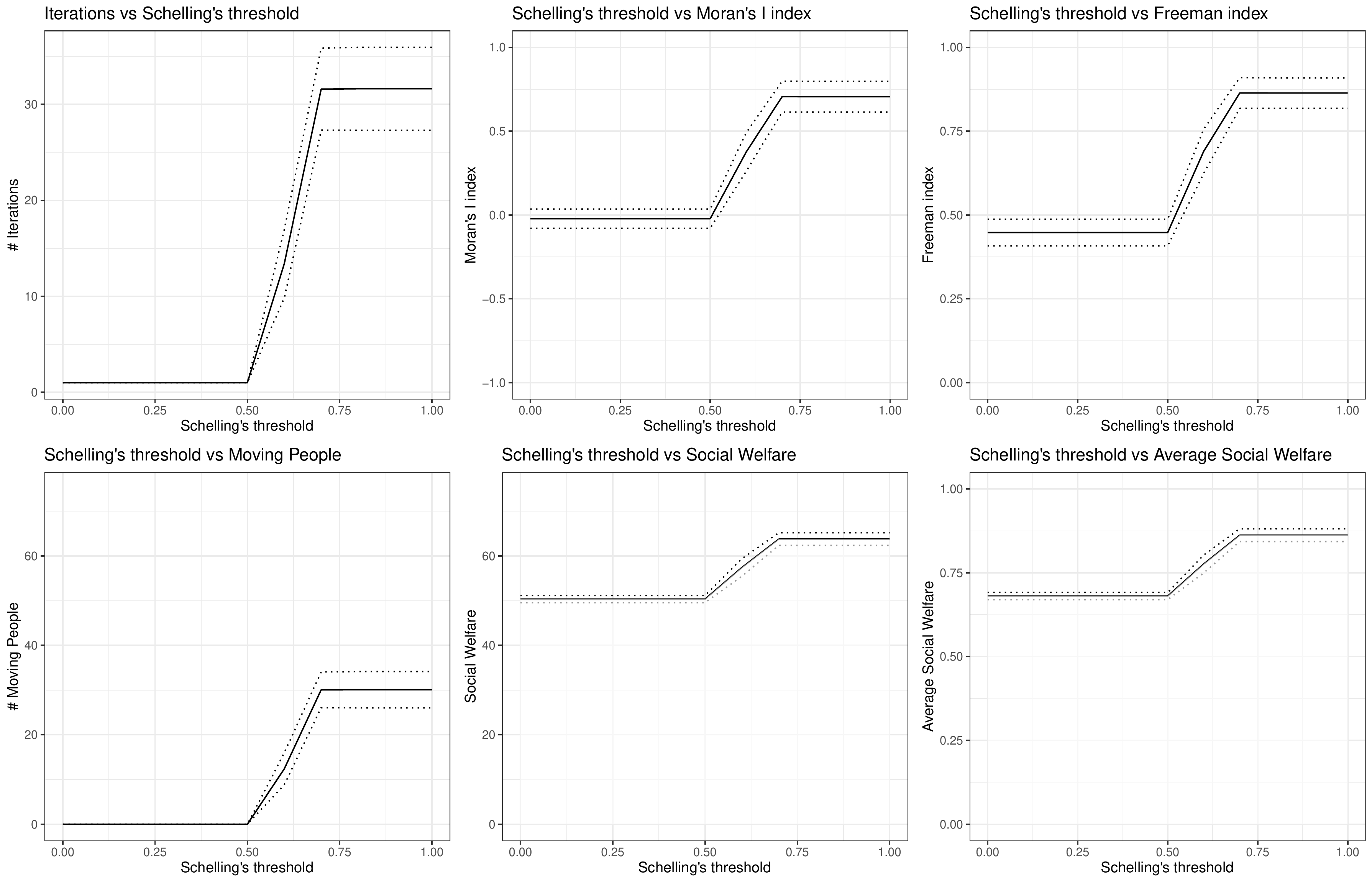}
        \caption[caption]{\scriptsize{No Moving costs ($\beta = 1$) - No utility from friendship ($\alpha = 1$).\\\hspace{\textwidth}Each iteration is repeated 100 times by permuting the initial torus. On the abscisses, the Schelling's threshold $x$ ranging in $[0,1].$ Bold and ticked lines indicate respectively average value and standard deviation. Upper left panel indicates the number of iterations required for the model to converge for different $x$ values. Upper middle and right panels report respectively the Moran's I and the FSI value obtained after model convergence. Bottom left panel shows the number of people who moved at least once in a simulation. Bottom middle and right panels display respectively the average and the total social welfare of the agents at the end of the iteration process. Social welfare is obtained from the component $\beta \alpha U_i^{color}(\bar{v};x)$ of the $i$-th agent utility function (equation (\ref{UF})).}}
        \label{fig:threshold}
    \end{center}
\end{figure}

This first simulation experiment sets the baseline to evaluate the other results. Results are shown in Figure \ref{fig:threshold}, which illustrates the change in each outcome (on the $y$-axes) for increasing levels of the Schelling's threshold $x$.

Model results are straightforward and consistent with the original Schelling model. When $x = 0$ (and, we stress it, agents assign no value to social connections: i.e., $\alpha = 1$), homophily play no rule in agents' relocation choices. Agents are indifferent to the density of same-color people in the neighborhood, and they have no incentives to move. As a result, the initial configuration remains unchanged and social welfare is unaltered. By contrast, when there is a preference for neighborhoods characterized by a majority of same-color agents, i.e. $x \geq 0.5$, people begin to move, increasing the level of both segregation and social welfare. Specifically, we observe that when $x>0.7$, total segregation is almost reached, and approximately 50\% of the population is involved in a change of location.

\subsection{The Schelling's model with the moving costs}\label{sec:movingcosts}
\begin{figure}[H]
	\begin{center}
		\includegraphics[width=0.7\textwidth]{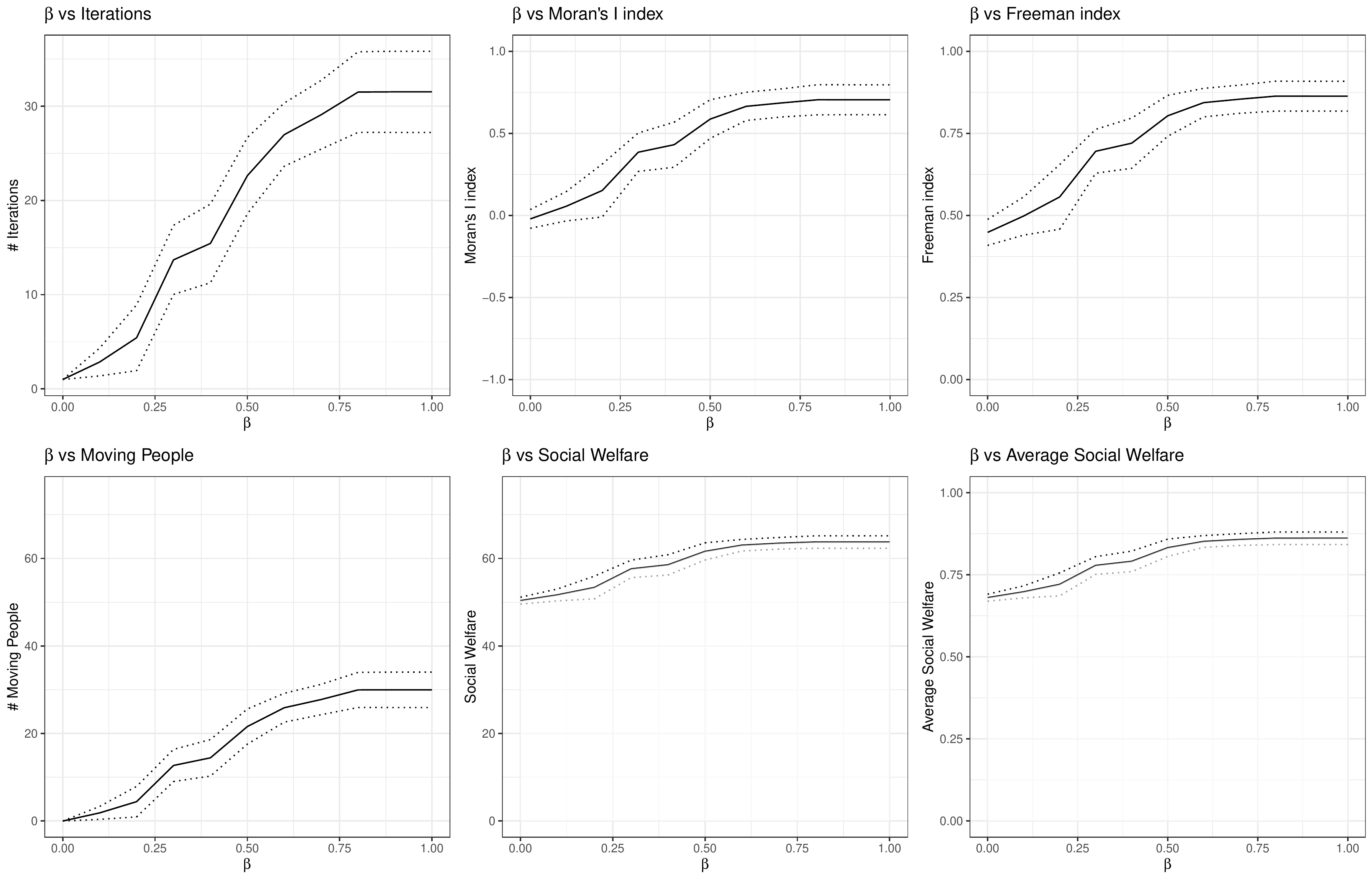}
		\caption[caption]{
\scriptsize{Fixed fair moving costs ($\gamma = 1,\ \bar{c} = 0.5$) - Maximum value of the Schelling's threshold $x=1$ - No contribution of friendship to utility ($\alpha = 1$).\\\hspace{\textwidth} On the abscisses, the parameter $\beta$ ranging in $[0,1]$. Each iteration is repeated 100 times by permuting the initial torus. Bold and ticked lines indicate respectively average value and standard deviation. Upper left panel indicates the number of iterations required for the model to converge. Upper middle and right panels report respectively the Moran's I and the FSI value obtained after model convergence. Bottom left panel shows the number of people who moved at least once in a simulation. Bottom middle and right panels display respectively the average and the total social welfare of the agents at the end of the iteration process. Social welfare is obtained from the component $\beta \alpha U_i^{color}(\bar{v};x)$ of $i$-th agent's utility function (equation (\ref{UF})).}}
		\label{fig:increasing_05_fc}
	\end{center}
\end{figure}

\begin{figure}[H]
	\begin{center}
		\includegraphics[width=0.7\textwidth]{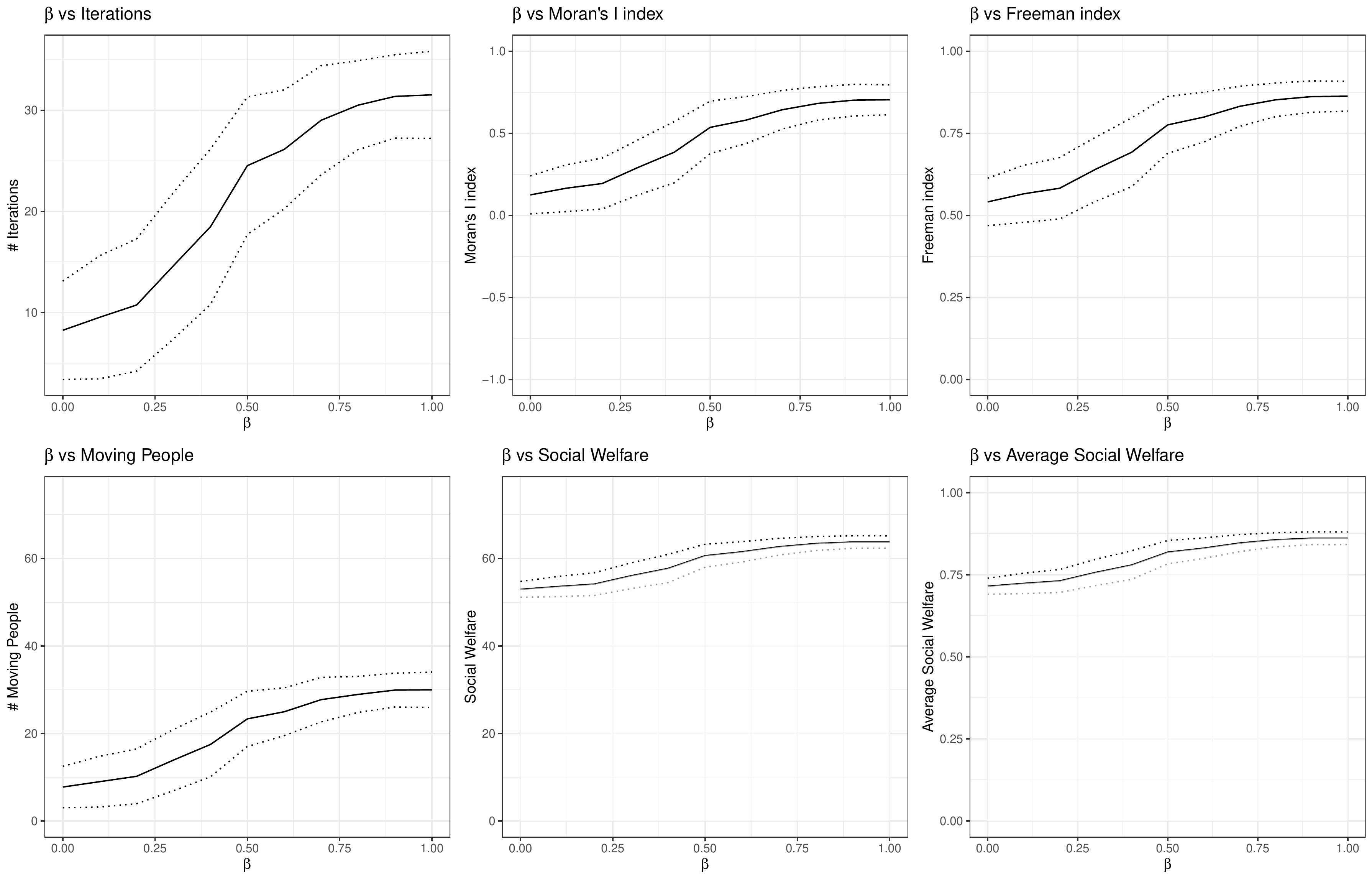}
		\caption[caption]{\scriptsize{Variable fair moving costs ($\gamma = 0,\ \bar{c} = 0.5$) - Maximum value of the Schelling's threshold $x=1$ - No contribution of friendship to utility ($\alpha = 1$).\\\hspace{\textwidth} On the abscisses, the parameter $\beta$ ranging in $[0,1]$. Each iteration is repeated 100 times by permuting the initial torus. Bold and ticked lines indicate respectively average value and standard deviation. Upper left panel indicates the number of iterations required for the model to converge. Upper middle and right panels report respectively the Moran's I and the FSI value obtained after model convergence. Bottom left panel shows the number of people who moved at least once in a simulation. Bottom middle and right panels display respectively the average and the total social welfare of the agents at the end of the iteration process. Social welfare is obtained from the component $\beta \alpha U_i^{color}(\bar{v};x)$ of $i$-th agent's utility function (equation (\ref{UF})).}}
		\label{fig:increasing_05_vc}
	\end{center}
\end{figure}

The role played by moving costs is to affect agents' location choices by mitigating the attraction force determined by their preference over homophily. In particular, this component of the agent's utility function discourages long movements and contrast the emergence of complete segregation. In order to show this effect, we simulate how the model initial configuration is changed for different levels of agents' responsiveness to either fixed or variable moving costs, i.e. with respect to $\beta$ in $[0,1]$. Specifically, we set $x=1$, $\alpha=1$, $\bar{c}=0.5$ and the two extreme scenarios $\gamma=0$ and $\gamma=1$. The results of this exercise are presented respectively in Figure \ref{fig:increasing_05_fc}, where we impose a fixed cost on agents' moves, and in Figure \ref{fig:increasing_05_vc}, where we allow agent's moving costs to be variable. In both figures, the plots show the effect of a change in  $\beta$, displayed on the x-axes, on one of the six outcomes considered, reported on the y-axis.

Consistent with expectations, the results show a monotonic relation between agents' responsiveness and the number of individual movements: i.e., the lower the moving costs, the higher is the number agents changing their location (bottom-left plots). Analogously, when moving costs decreases, the number of model iterations increases, and convergence timing is delayed (top-left plots).

In addition, we see that when agents are left free to relocate (i.e. $\beta$ decreases), the model tends to mimic Schelling's results: i.e., blue and red nodes move in different and non-contiguous areas, and segregation increases in terms of both $FSI$ and Moran's $I$ values (top-middle and right panel). This is not surprising, since network effects are null (i.e. degree centrality is equal to 0), and Schelling's heuristic is the only determinant of movements.

Finally, we witness a negative relation between moving costs and social welfare (bottom-middle and right panel). Specifically, the highest levels of agents satisfaction are reached when $\beta \geq 0.75$. Above this value, $U_{i}(\bar{v})$ remains somewhat constant.

The robustness of these results is confirmed by further analyses presented in the Appendices, where we show that results are left qualitatively unchanged when replicating this exercise using low fixed costs (Figure \ref{fig:increasing_001_fc}, with $\gamma=1$ and $\bar{c}=0.01$), low variable costs (Figure \ref{fig:increasing_001_vc}, with $\gamma=0$ and $\bar{c}=0.01$), high fixed costs (Figure \ref{fig:increasing_099_fc}, with $\gamma=1$ and $\bar{c}=0.99$), and high variable costs (Figure \ref{fig:increasing_099_vc}, , with $\gamma=1$ and $\bar{c}=0.99$).

\subsection{The Schelling's model with the moving costs and the networked individuals}\label{sec:networked}

In the third experiment, when $\alpha$ is different from 1, agents' degree centrality $k$ is taken equals to 3, to highlight the difference between no friends and some friends. The scenarios for $\alpha$ are $\alpha=1$ (utility does not depend on the Schelling's heuristics) and $\alpha=0.5$, which means that agents assign equal weights to neighborhood composition and friends' distance. The value of $\beta$ is set to 0.5, to represent a fair introduction of the moving costs in the utility. The impact of this parameter on the model outcome is alternatively tested by setting $\gamma=0$ (i.e. fixed costs) and $\gamma=1$ (i.e. variable costs). We also investigate changes in the model outcome when moving costs are removed by taking $\beta=1$. The simulation experiments are performed by taking into account the sensitivity with respect to Schelling's threshold $x$.

\begin{figure}[H]
    \begin{center}
        \includegraphics[width=0.65\textwidth]{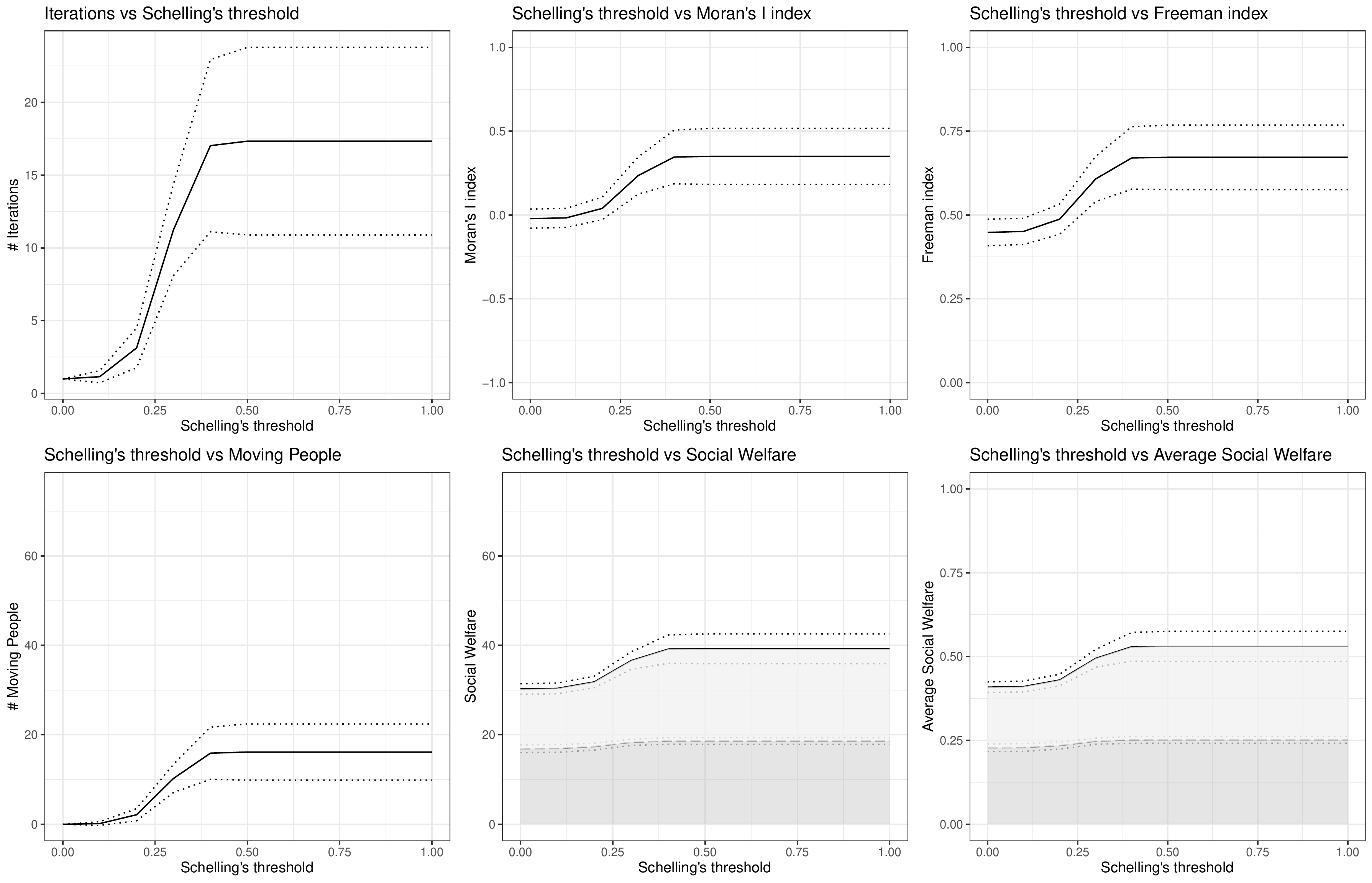}
        \caption[caption]{\scriptsize{No Moving costs ($\beta = 1$) - fair utility from friendship ($\alpha=0.5$) - 3 friends ($k = 3$)\\\hspace{\textwidth}Each iteration is repeated 100 times by permuting the initial torus and network connections. On the abscisses, the Schelling's threshold $x$ ranging in $[0,1].$ Bold and ticked lines indicate respectively average value and standard deviation. Upper left panel indicates the number of iterations required for the model to converge. Upper middle and right panels report respectively the Moran's I and the FSI value obtained after model convergence. Bottom left panel shows the number of people who moved at least once in a simulation. Bottom middle and right panels display respectively the average and the total social welfare of the agents at the end of the iteration process. Social welfare is obtained from $i$-th agent's utility function (equation (\ref{UF})). The color of the area indicates the two components of agent's utility after convergence is achieved: $\beta \alpha U_i^{color}(\bar{v};x)$ (lighter-colored area) and $\beta \alpha U_i^{friend}(\bar{v})$ (darker-colored area).}}
        \label{fig:threshold_friends}
    \end{center}
\end{figure}

We first refer to results in Figure \ref{fig:threshold_friends}. The introduction of the friendship network has a dramatic impact on the results. The force of attractiveness generated by the homophily preferences is now bonded by the different characteristics with which agents identify (i.e. the color and the social connection). Moreover, agents' movements are constrained by the tension exerted by friends' position, who are in turn attracted by their friends. We first focus on this latter element. When neighborhood composition is irrelevant, i.e. $x=0$, friendship drives homophily and generates two competing forces. The first is an attractive force, like a spring, which brings agents that are linked closer together. The second is a repulsive force, generated by the presence of multiple connections, which prevents long movements of people to the same area, and forces friends apart from each other. Put differently, network externalities are at play, and their effects cascade quite dramatically. As a result, agents act as if they were steel rings linked by springs, and any radical change in the original status quo is prevented: i.e., less than 25\% of the population choose to change location and model convergence is reached with few iterations. It follows that total segregation is far from being reached in this setting.

However, when agents begin to care about the characteristics of their neighbors,  $x>0$, a significant change occurs on their preferences over the composition of their neighbors. This results in a reduction of the local inertia determined by network externalities, and the set of individual relocation choices expands. This produces an increase of both segregation and social welfare, but interestingly, both outcomes remain significantly lower than those achieved in Figure \ref{fig:threshold}, meaning that the introduction of the network has the simultaneous effect of reducing segregation and producing sub-optimal levels of welfare.

This finding provides an interesting insight. When homophily requires to combine multiple characteristics, agents find it difficult to find a neighborhood that improves her levels of welfare: e.g. a racially mixed neighborhood where the majority is composed by people of her own color, and the minority is represented by her social connections.

\begin{figure}[H]
	\begin{center}
		\includegraphics[width=0.65\textwidth]{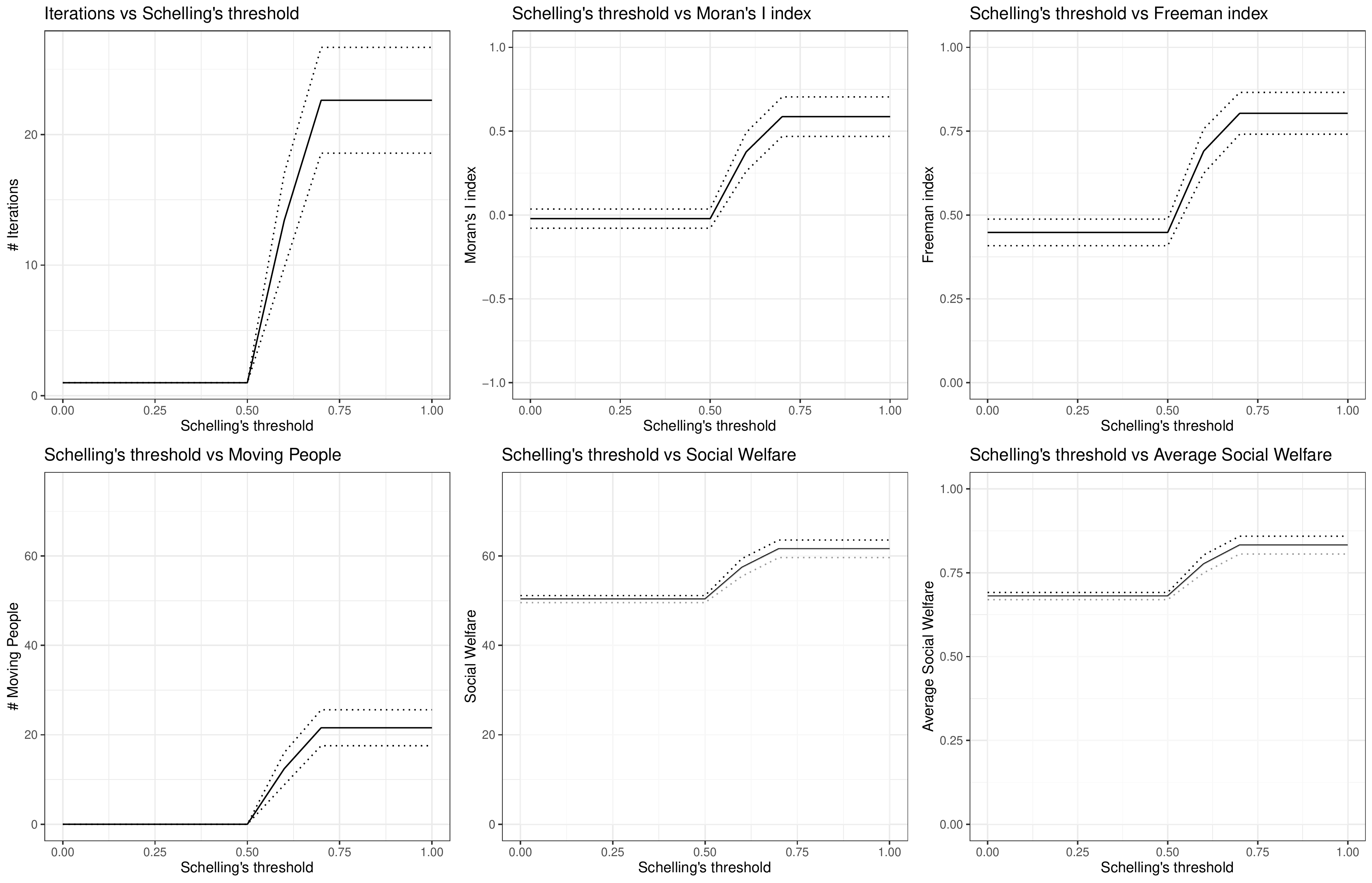}
		\caption[caption]{\scriptsize{
				Fixed moving costs ($\beta = 0.5,\ \gamma = 1,\ \bar{c} = 0.5$) - no utility from friendship ($\alpha = 0$).\\\hspace{\textwidth}Each iteration is repeated 100 times by permuting the initial torus. On the abscisses, the Schelling's threshold $x$ ranging in $[0,1].$ Bold and ticked lines indicate respectively average value and standard deviation. Upper left panel indicates the number of iterations required for the model to converge. Upper middle and right panels report respectively the Moran's I and the FSI value obtained after model convergence. Bottom left panel shows the number of people who moved at least once in a simulation. Bottom middle and right panels display respectively the average and the total social welfare of the agents at the end of the iteration process. Social welfare is obtained from $i$-th agent's utility function (equation (\ref{UF})).}}
		\label{fig:threshold_05_fc}
	\end{center}
\end{figure}

\begin{figure}[H]
    \begin{center}
        \includegraphics[width=0.65\textwidth]{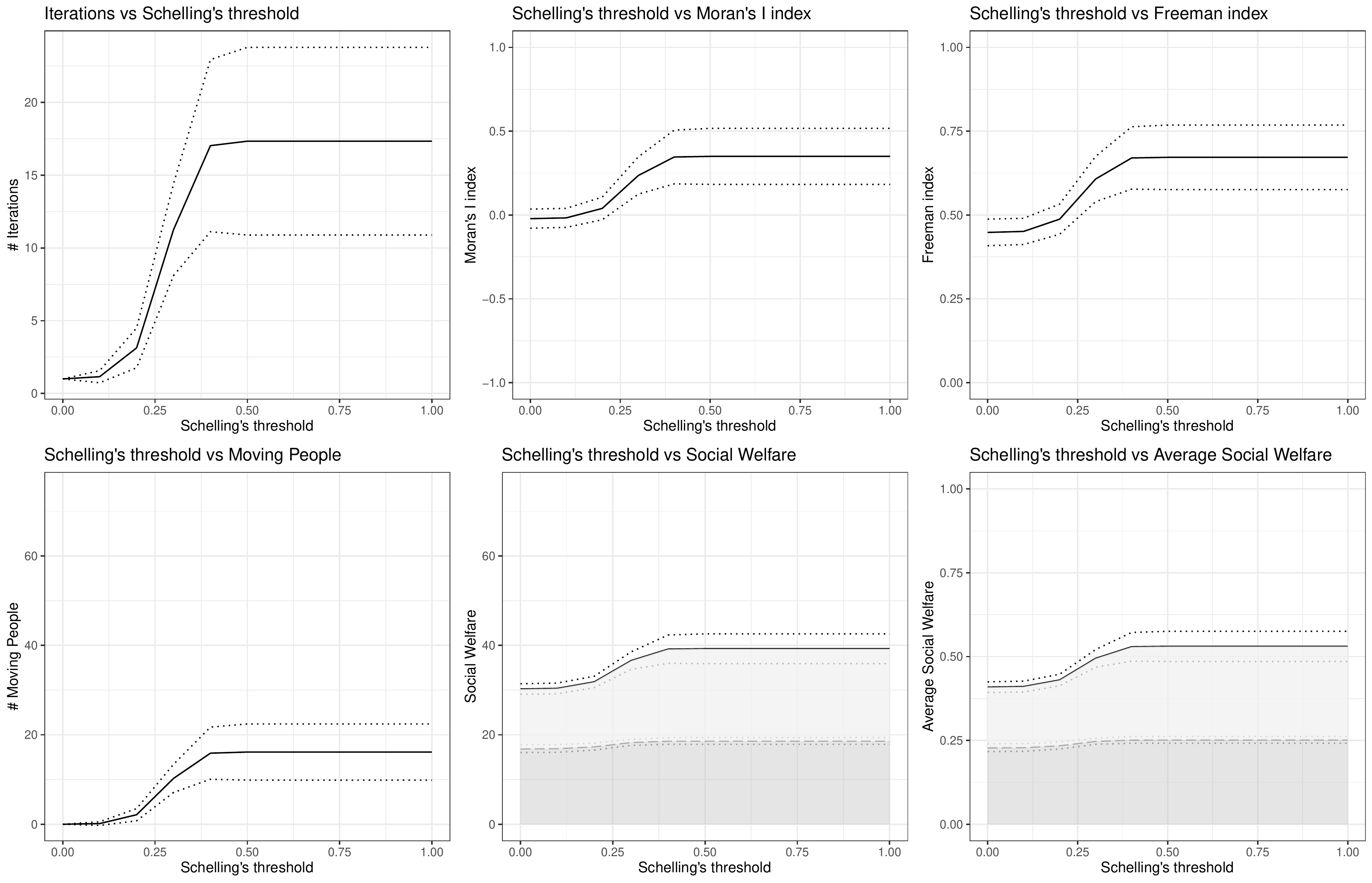}
        \caption[caption]{\scriptsize{Fixed moving costs ($\beta = 0.5,\ \gamma = 1,\ \bar{c} = 0.5$) - fair dependence of the utility on friendship ($\alpha = 0.5$) - 3 friends ($k=3$).\\\hspace{\textwidth}Each iteration is repeated 100 times by permuting the initial torus. On the abscisses, the Schelling's threshold $x$ ranging in $[0,1].$ Bold and ticked lines indicate respectively average value and standard deviation. Upper left panel indicates the number of iterations required for the model to converge. Upper middle and right panels report respectively the Moran's I and the FSI value obtained after model convergence. Bottom left panel shows the number of people who moved at least once in a simulation. Bottom middle and right panels display respectively the average and the total social welfare of the agents at the end of the iteration process. Social welfare is obtained from $i$-th agent's utility function (equation (\ref{UF})). The color of the area indicates the two components of agents' utility after convergence is achieved: $\beta \alpha U_i^{color}(\bar{v};x)$ (lighter-colored area) and $\beta \alpha U_i^{friend}(\bar{v})$ (darker-colored area).}}
        \label{fig:threshold_friends_05fc}
    \end{center}
\end{figure}

\begin{figure}[H]
    \begin{center}
        \includegraphics[width=0.65\textwidth]{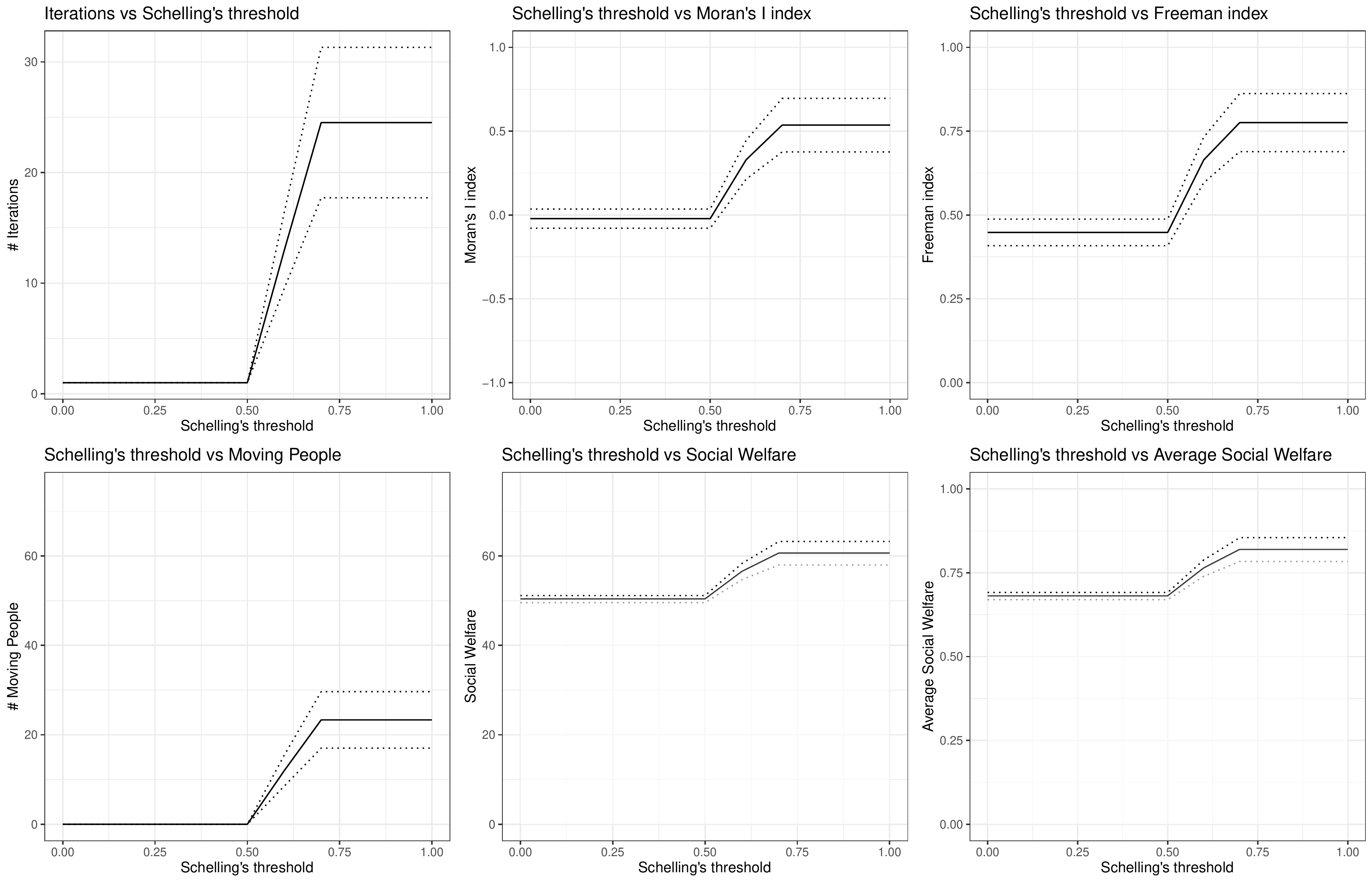}
        \caption[caption]{\scriptsize{
                Variable moving costs ($\beta = 0.5,\ \gamma = 0,\ \bar{c} = 0.5$) - No utility from friendship ($\alpha=1$).\\\hspace{\textwidth}Each iteration is repeated 100 times by permuting the initial torus. On the abscisses, the Schelling's threshold $x$ ranging in $[0,1].$ Bold and ticked lines indicate respectively average value and standard deviation. Upper left panel indicates the number of iterations required for the model to converge. Upper middle and right panels report respectively the Moran's I and the FSI value obtained after model convergence. Bottom left panel shows the number of people who moved at least once in a simulation. Bottom middle and right panels display respectively the average and the total social welfare of the agents at the end of the iteration process. Social welfare is obtained from $i$-th agent's utility function (equation (\ref{UF})).}}
        \label{fig:threshold_05_vc}
    \end{center}
\end{figure}

\begin{figure}[H]
    \begin{center}
        \includegraphics[width=0.65\textwidth]{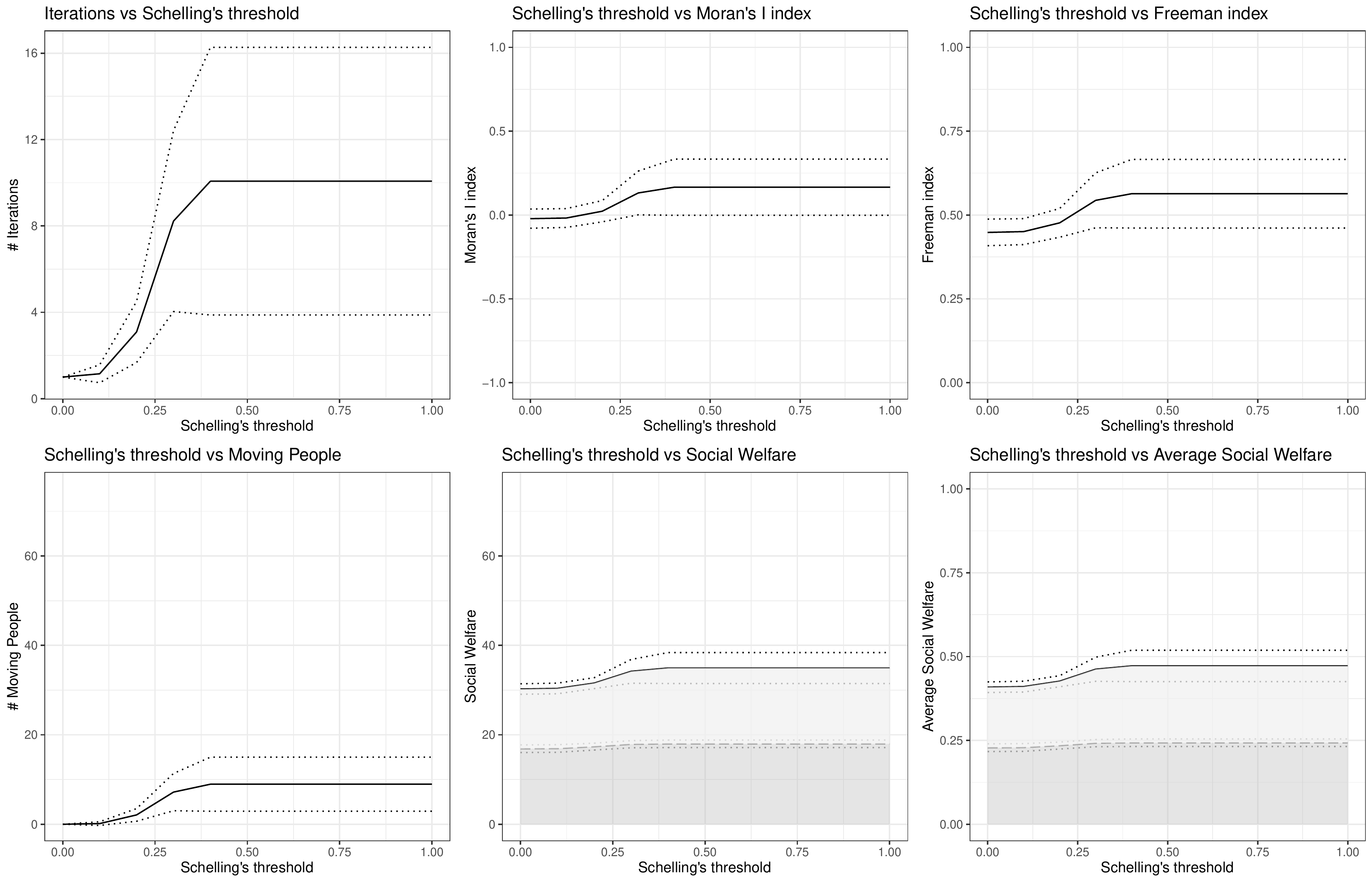}
        \caption[caption]{\scriptsize{Variable moving costs ($\beta = 0.5,\ \gamma = 0,\ \bar{c} = 0.5$) - Fair utility from friendship and 3 friends ($\alpha=0.5, \ k = 3$).\\\hspace{\textwidth}Each iteration is repeated 100 times by permuting the initial torus. On the abscisses, the Schelling's threshold $x$ ranging in $[0,1].$ Bold and ticked lines indicate respectively average value and standard deviation. Upper left panel indicates the number of iterations required for the model to converge. Upper middle and right panels report respectively the Moran's I and the FSI value obtained after model convergence. Bottom left panel shows the number of people who moved at least once in a simulation. Bottom middle and right panels display respectively the average and the total social welfare of the agents at the end of the iteration process. Social welfare is obtained from $i$-th agent's utility function (equation (\ref{UF})). The color of the area indicates the two components of agents' utility after convergence is achieved: $\beta \alpha U_i^{color}(\bar{v};x)$ (lighter-colored area) and $\beta \alpha U_i^{friend}(\bar{v})$ (darker-colored area).}}
        \label{fig:threshold_friends_05vc}
    \end{center}
\end{figure}

When we set $\beta = 0.5$, then we observe a contribution to our analysis, regardless of whether we consider network effects or not. In fact, significant changes are registered neither when costs are fixed (Figures \ref{fig:threshold_05_fc} and \ref{fig:threshold_friends_05fc}), nor when costs are variable (Figures \ref{fig:threshold_05_vc} and \ref{fig:threshold_friends_05vc}), and results remain qualitatively the same.\footnote{Additional evidence is reported in \ref{sec:appendix.figures.low.cost}, where we test the effect of reducing fixed costs ($\gamma=1$ and $\bar{c}=0.01$, see Figures \ref{fig:threshold_001_fc} and \ref{fig:threshold_friends_001_fc}) and variable costs ($\gamma=0$ and $\bar{c}=0.01$, see Figures \ref{fig:threshold_001_vc} and \ref{fig:threshold_friends_001_vc}). Similarly, in \ref{sec:appendix.figures.high.cost}, we test the effect of increasing fixed costs ($\gamma=1$ and $\bar{c}=0.99$, see Figures \ref{fig:threshold_099_fc} and \ref{fig:threshold_friends_099_fc}) and variable costs ($\gamma=1$ and $\bar{c}=0.99$, see Figures \ref{fig:threshold_099_vc} and \ref{fig:threshold_friends_099_vc}).} All in all, the introduction of moving costs only produces a shift in the outcomes value.

The only difference found is when we consider the combination of high fixed moving costs and no friends ($\gamma=1$, $\alpha=1$ and $\bar{c}=0.99$, see Figure \ref{fig:threshold_099_fc}). In fact, in this case there is no incentive to move, convergence is achieved after 1 iteration and the original status quo remains unchanged. In other words, we observe a different mechanism for the configuration of a system of local inertia, where the effect of homophily is nullified by moving costs this time. It should not be surprising that the same does not apply to the case of variable costs ($\gamma=0$, $\alpha=1$ and $\bar{c}=0.99$, see Figure \ref{fig:threshold_099_vc}). While in the former case any movement has the same cost, in the latter setting the cost of changing location is mitigated by distance, and short movements are still possible.

\subsection{An extreme case of Schelling's model and the networked individuals}
\begin{figure}[H]
    \begin{center}
        \includegraphics[width=0.65\textwidth]{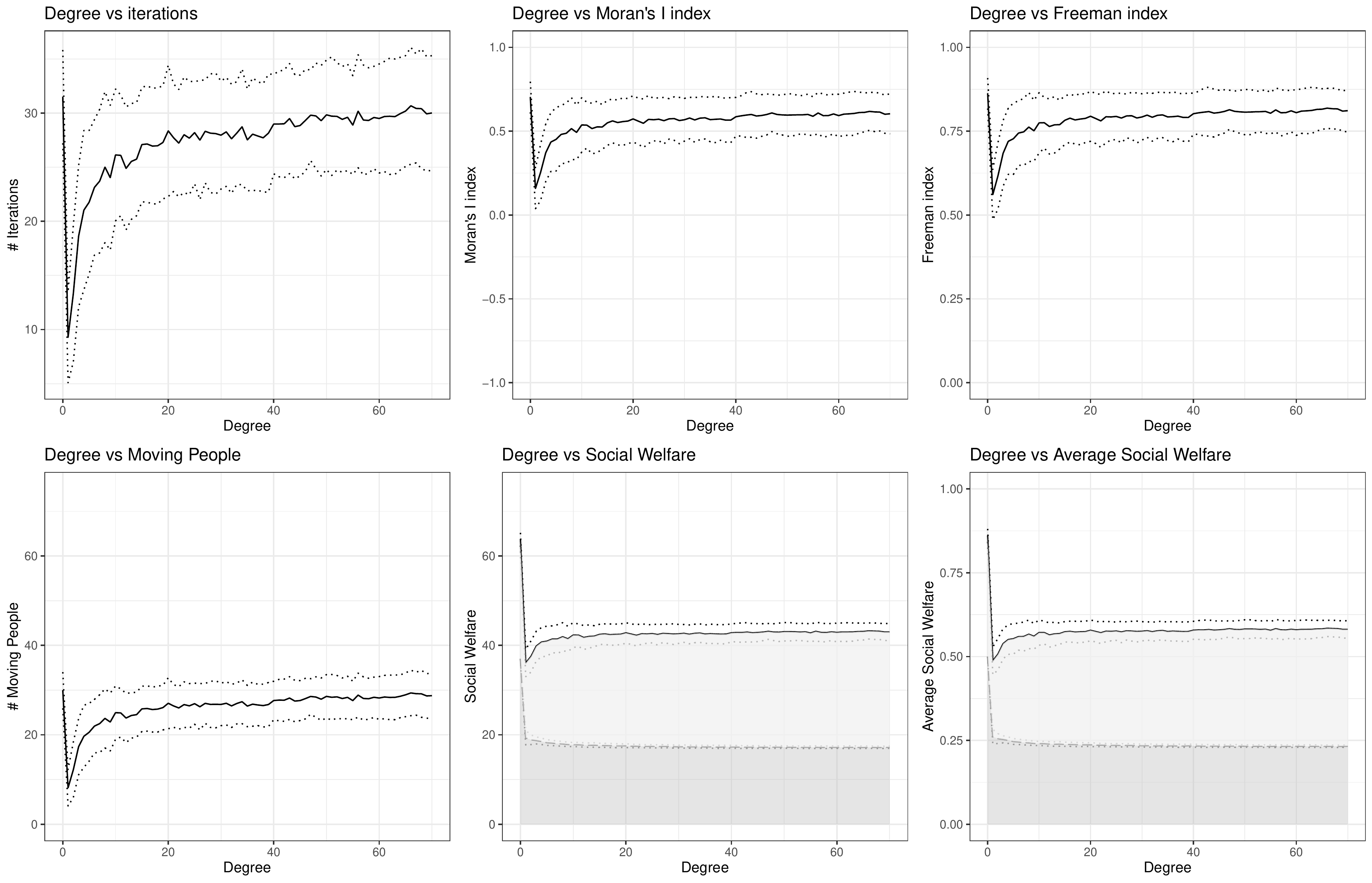}
        \caption[caption]{\scriptsize{No moving costs ($\beta = 1$) - Fair contribution of friendship to utility ($\alpha=0.5$) - Maximum value of the Schelling's threshold $x=1$\\\hspace{\textwidth}Each iteration is repeated 100 times by permuting the initial torus and network connections. On the abscisses, the degree centrality (number of friends) $k$, ranging from 0 to 73. Bold and ticked lines indicate respectively average value and standard deviation. Upper left panel indicates the number of iterations required for the model to converge. Upper middle and right panels report respectively the Moran's I and the FSI value obtained after model convergence. Bottom left panel shows the number of people who moved at least once in a simulation. Bottom middle and right panels display respectively the average and the total social welfare of the agents at the end of the iteration process. Social welfare is obtained from agents' utility function (equation (\ref{UF})). The color of the area indicates the two components of agents' utility after convergence is achieved: $\beta \alpha U_i^{color}(\bar{v};x)$ (lighter-colored area) and $\beta \alpha U_i^{friend}(\bar{v})$ (darker-colored area).}}
        \label{fig:increasing_friends}
    \end{center}
\end{figure}

\begin{sidewaysfigure}
	\begin{figure}[H]
		\begin{center}
			\centerline{  \includegraphics[width=1.1\textwidth]{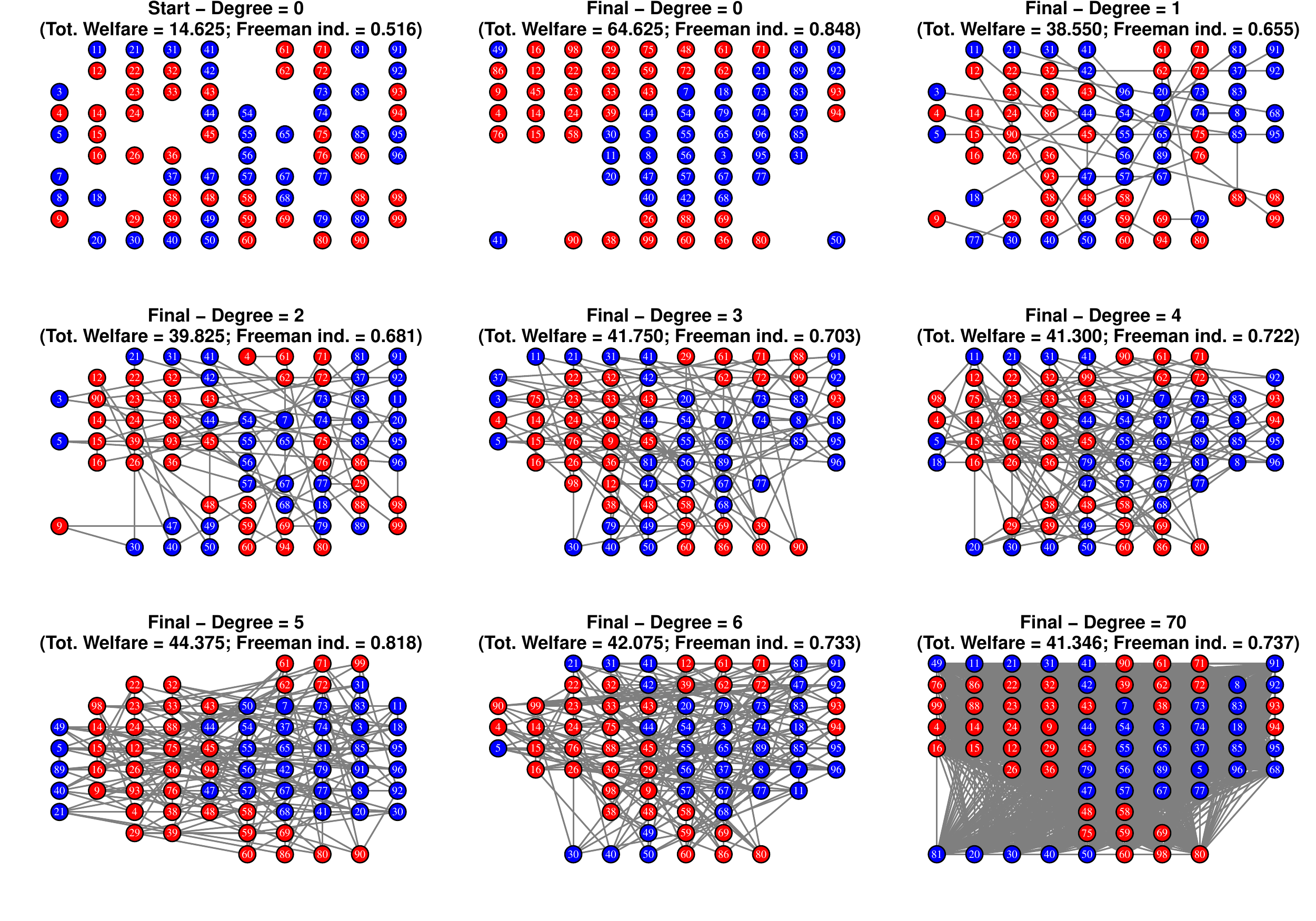}}
			\caption[caption]{\scriptsize{The Figure on the upper-left panel shows the starting position of the torus, and it indicates the values of total welfare and Freeman index when degree is equal to 0. Other figures shows the final position of the torus corresponding to different levels of degree. Also for this cases, we report the values of total welfare and Freeman index.}}  \label{fig:network}
		\end{center}
		\vspace{-10mm}
	\end{figure}
\end{sidewaysfigure}

The effect of the network on agents' location choice when neighborhoods composition matters is further investigated in this section. In the experiment summarized in Figure \ref{fig:increasing_friends}, Schelling's threshold is stressed by setting $x$ equals to $1$, meaning that agents are sensitive to any marginal change in their neighborhood composition. Moving costs are removed, by taking $\beta=1$, while $\alpha=0.5$, which means that there is a fair contribution of Schelling's heuristics and friendships. Degree centrality $k$ is let to vary iteratively, from 0 to 73. Therefore, now the abscisses of the plots reports the level of degree centrality, while the y-axes are used once again to register changes in the outcome values. For illustrative purposes, Figure \ref{fig:network} shows also various instances of the final torus configuration when increasing degree centrality.

As expected, when degree centrality is 0, the model replicates the last observation registered in the plots of Figure \ref{fig:threshold}. Then, when agents form a friendship, a significant change occurs in the formation of individual homophily preferences. As a result, the utility exerted from positioning close to an agent is now equivalent to the utility obtained by locating in a neighborhood where all agents have the same color. In other words, we observe the configuration of a peculiar physical system unconstrained by the repulsive forces generated by the composition of the neighborhood or by multiple friendships. If agents choose a location close to their friend, this in turn has no need to relocate. Consequently, the utility obtained by the composition of the neighborhood suddenly drops, and we observe a significant decrease in the segregation indices.

The externalities produced by the combination of Schelling's heuristic and social linkages are an important component of our model, because they show how competing identities modify homophily effects. As agents reach a neighborhood, others who are in the neighborhood might decide to remain regardless of their color, and the composition of their neighbors. This mechanism generates limited segregation equilibria and suggests an explanation for the existence of racially mixed neighborhoods. This is somewhat the case described in the upper-right panel in Figure (\ref{fig:network}): agent 69 and 79 are friends living in adjacent cells, so even if the latter does not enjoy the composition of the neighborhood, they decide not to move. A similar behavior is observed for agents 18 and 36. Since 36 benefits from the new composition of the neighborhood, and she has not incentives to move, 18 decides to remain in the starting position, which is relatively close to 36, even if no blue agents are in the surroundings.

However, when $k$ increases, agents begin to look for an equilibrium in between the distance from their friends and the neighborhood where they live in. In this attempt to manage multiple preferences, agents seem to give more importance to the composition of their neighborhood than to their friends location, since the utility obtained by the former factor increases, while the latter decreases. This might be due to the fact that agents find easier to locate in a neighborhood with specific characteristics and close to few friends, than finding a place that is suitable for all her friends. This is for example the case of agents 9 and 36, in the middle-left panel of Figure \ref{fig:network}. Both of them are relatively close to one of their friends, and located in a neighborhood where all agents share her own color.

This is another important finding of this study. Competing characteristics do not equally matter in forming agents homophily preferences. Rather, certain characteristics featured by the agent tend to prevail in the definition of her own identity because it is easier to locate close to someone with that same characteristics (e.g. color over social linkages). Hence, even when agents assign the same importance to all their characteristics (e.g. $\alpha = 0.5$), some of them will have a stronger impact on relocation choices.

This mechanism is even clearer if we consider the case where each individual is connected with more than a relatively small proportion of the population (c.a. 5-8\%). In fact, when the number of friendships becomes too high, there is an increasing disutility in locating close to few friends: e.g. when degree centrality is 73, even if $i$ is surrounded by all friends of her color, she is still far from 65 friends. This implies that as the degree increases, the location of friends becomes somewhat indifferent to agents, because for each friend found in one place, there will be many others left far apart, and most of the utility is obtained by the composition of the neighborhood: this is for example the case of agent 44, who despite the increase in her degree centrality, she remained consistently in her starting locations, next to 3 blue nodes, in 

all panels of Figure (\ref{fig:network}).

Moreover, it is worth stressing that because $\alpha = 0.5$, competing characteristics counteract each other, even if one of them is producing a disutility (e.g. social connections). Put differently, agents are able to cope with the disutility derived from one of their characteristics and maintain social welfare constant by letting Schelling's heuristic prevail over other utility's components in their relocation choices.

\section{Discussion and Conclusions}

This paper presented an extension of the original racial and residential segregation model by Schelling, i.e. ``chequerboard model''. Specifically, the aim was to investigate whether and how model's predictions can be improved in order to incorporate solutions of limited segregation equilibria.

The evidence provided by our simulations suggests that a possible explanation for the formation of racially mixed areas might be the presence of moving costs and the effect of social linkages. Moving costs, regardless of whether they are fixed or variable, consistently determines phenomena of local inertia, preventing agents to leave their neighborhood for another that better suits with their set of preferences. Interestingly, a similar outcome is observed when agents have a stronger preference for social linkages over Schelling's heuristic. In this case, agents' movements are constrained by the tension exerted by friends' position, who are in turn constrained in their movements by their friends. Perhaps unexpectedly, local inertia is found also to be associated with lower levels of social welfare.

Another interesting insight provided by our study is that, whenever homophily effects are constrained by the existence of competing characteristics with which agents equally identify (e.g. ethnicity and social connections), certain characteristics tend to prevail in the formation relocation choices. The reason is straightforward. It is easier for agents to identify a neighborhood where only one characteristic is predominant. Future research should be dedicated to understand what are these characteristics that tend to define one's own identity and drive relocation choice.

This investigation has also highlighted that whenever homophily requires to combine multiple characteristics, the generic agent finds it difficult to find a neighborhood that will improve her levels of welfare: e.g. a racially mixed neighborhood where the majority is composed by people of her own color, and the minority is represented by her social connections. This suggests that suboptimal levels of welfare might be endemic in context where multiple elements concur to the definition of one's own identity.

The very general setup of the model allows many interesting extensions that we leave for future research.\\
In the present model we kept Schelling's heuristic and the structure of the friendship network separated. This was done on purpose in order not to mix the effects of the two, but this modeling choice implies that the process shaping the racial distribution of friends is purely random and is not determined by a preference for homophily. Evidence is at odd. As recalled by \cite[p.109]{Jac2019}, simple accounting tells us that individuals in the minority group will end up having more friends from the majority group on average than individuals from the majority have with ones in the minority group. Considering homophily this effect would be magnified, rationalizing the striking evidence  reported by Jackson: ``Guess how many black friends the typical white person in the U.S. has - where a friend is someone with whom that person `regularly discussed important matters'? Zero.'' If friendship formation is lead by homophily, the two processes shaping Schelling's heuristic and friendship formation interact. A condition worth exploring in future research.

A second aspect that could add new insights to the model would be the possibility of forming or dissolving friendship at every movement on the chequerboard. Moving implies costs but also offers opportunity to create new friends and confirm the old ones. When social linkages varies with location changes, the new friends in the neighborhood reinforce the propensity not to further move. On the contrary, if past social linkages remain relevant, the incentive to go back to the previous location persists. This change in the model's setup would allow to study the role of past and present social linkages of migrants and the ultimate effect on segregation and welfare in the new location and in previous one.

Finally, the evidence of power laws, fat tails, Pareto and scale free degree distributions  is typical of social networks \citep{New2005, Goy09}. The model's assumption of a uniform distribution of friends is purely exemplificative and can be modified at no particular cost. In this case, centrality and degree heterogeneity would start to influence the segregation equilibria, since the network externality is associated to node centrality and the movement of central nodes brings a greater influence on the decision of poorly connected nodes. Similar extension can be applied to a social linkages that are directed and or weighted.


\newpage

\appendix
\counterwithin{figure}{section}
    \section{Figures: Low moving costs} \label{sec:appendix.figures.low.cost}
This appendix extends our analysis and presents additional results replicating the simulations in Section (\ref{sec:results}) with lower levels of moving costs ($\bar{c} = 0.01$) with respect to the main analysis ($\bar{c} = 0.5$). Please, see footnote 10 for the details of the selected parameter set.

\begin{figure}[H]
    	\begin{center}
    		\includegraphics[width=0.65\textwidth]{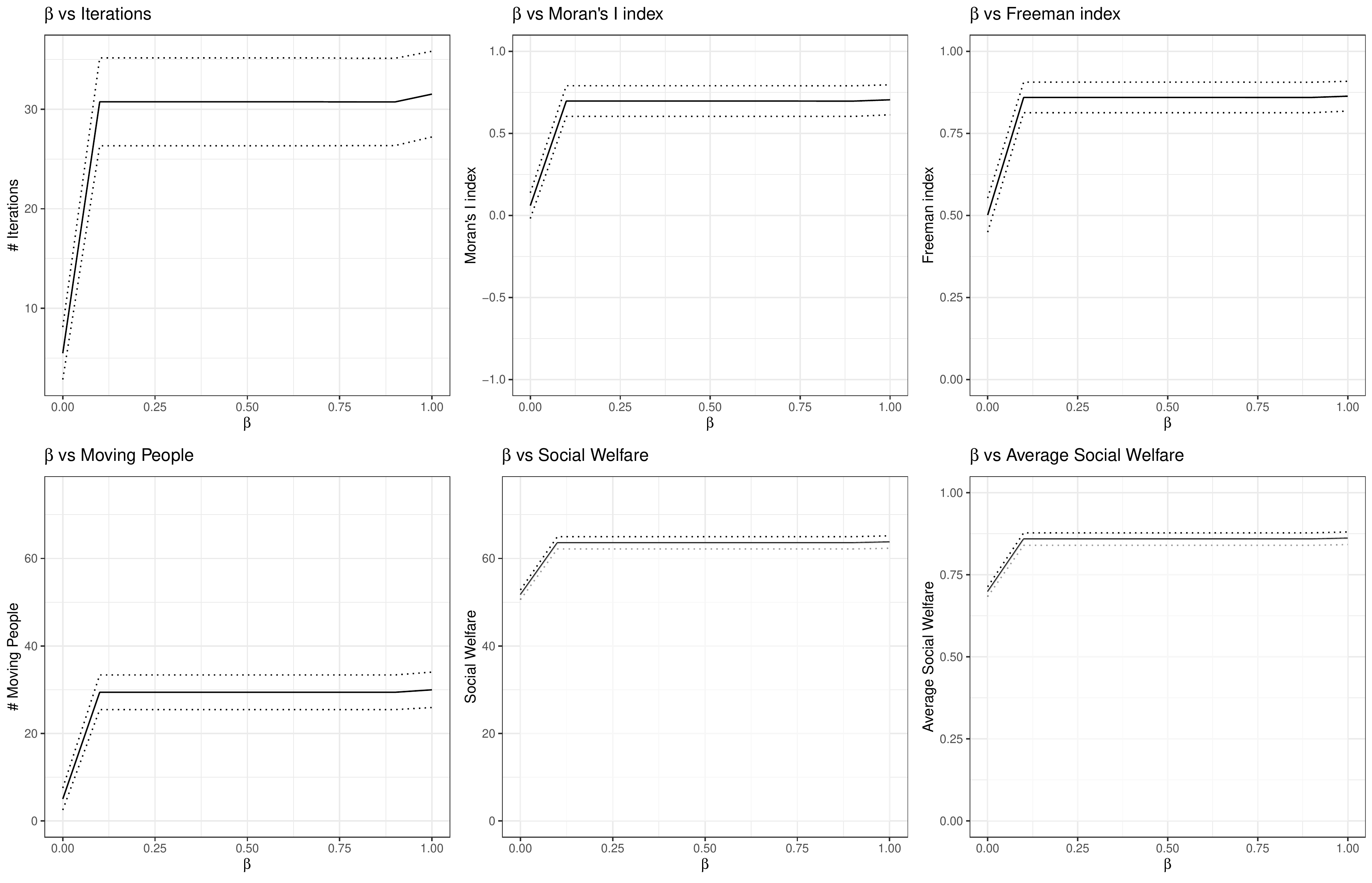}
    		\caption[caption]{\scriptsize{Low fixed moving costs ($\gamma = 1,\ \bar{c} = 0.01$) - Maximum value of the Schelling's threshold $x=1$ - No contribution of friendship to utility ($\alpha=1$).\\\hspace{\textwidth}Each iteration is repeated 100 times by permuting the initial torus. On the abscisses, we have parameter $\beta$ ranging in $[0,1]$.  Bold and ticked lines indicate respectively average value and standard deviation. Upper left panel indicates the number of iterations required for the model to converge. Upper middle and right panels report respectively the Moran's I and the FSI value obtained after model convergence. Bottom left panel shows the number of people who moved at least once in a simulation. Bottom middle and right panels display respectively the average and the total social welfare of the agents at the end of the iteration process. Social welfare is obtained from the component $\beta \alpha U_i^{color}(\bar{v};x)$ of $i$-th  agent utility function (equation (\ref{UF})).}}
    		\label{fig:increasing_001_fc}
    	\end{center}
    \end{figure}

\begin{figure}[H]
    	\begin{center}
    		\includegraphics[width=0.65\textwidth]{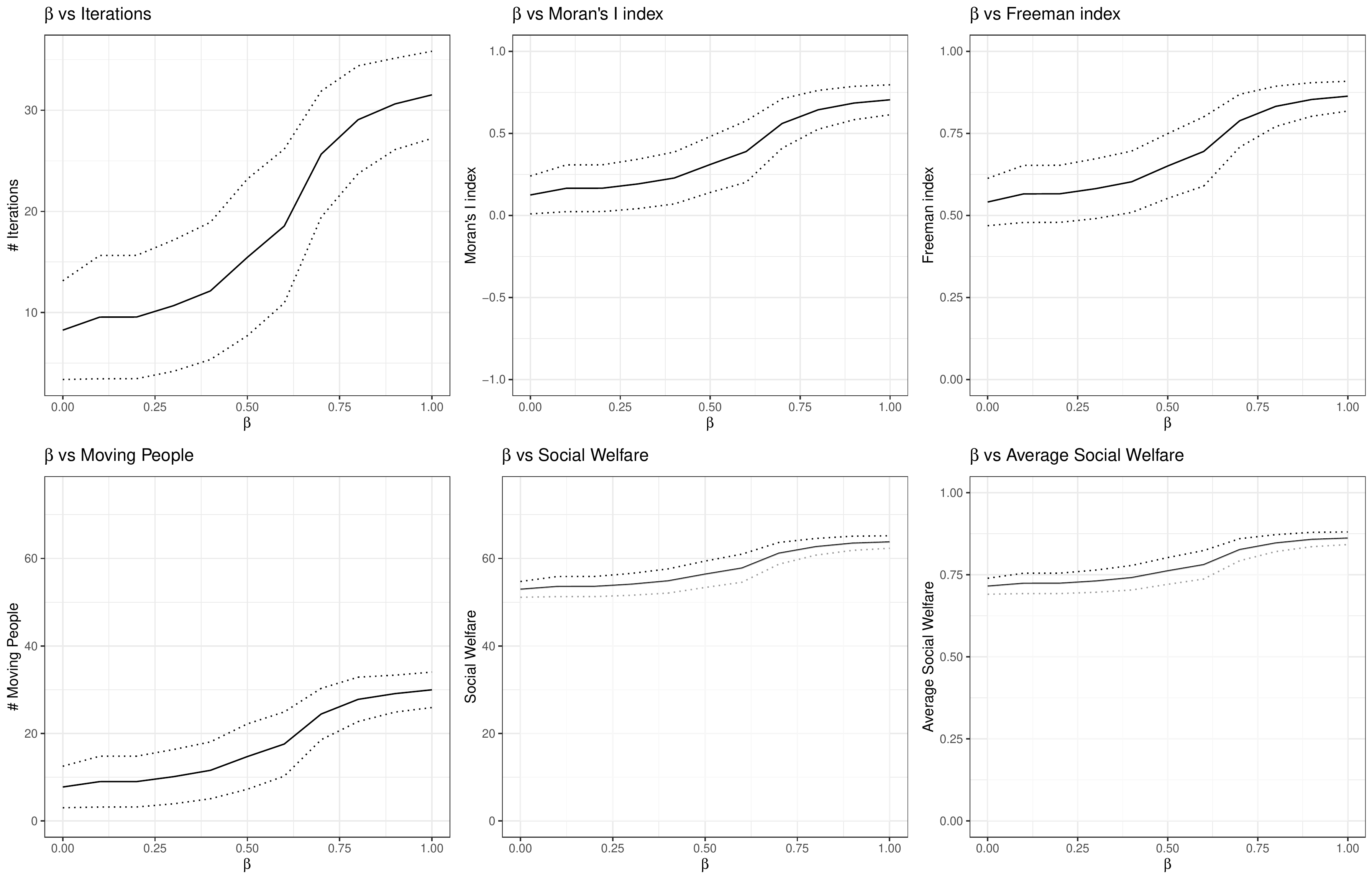}
    		\caption[caption]{\scriptsize{Low variable moving costs ($\gamma = 0,\ \bar{c} = 0.01$) - Maximum value of the Schelling's threshold $x=1$ - No contribution of friendship to utility ($\alpha=1$).\\\hspace{\textwidth}Each iteration is repeated 100 times by permuting the initial torus. On the abscisses, we have parameter $\beta$ ranging in $[0,1]$.  Bold and ticked lines indicate respectively average value and standard deviation. Upper left panel indicates the number of iterations required for the model to converge. Upper middle and right panels report respectively the Moran's I and the FSI value obtained after model convergence. Bottom left panel shows the number of people who moved at least once in a simulation. Bottom middle and right panels display respectively the average and the total social welfare of the agents at the end of the iteration process. Social welfare is obtained from the component $\beta \alpha U_i^{color}(\bar{v};x)$ of $i$-th  agent utility function (equation (\ref{UF})).}}
    		\label{fig:increasing_001_vc}
    	\end{center}
    \end{figure}

In Figures \ref{fig:increasing_001_fc} and \ref{fig:increasing_001_vc}, we progressively increase the responsiveness of $i$'s utility to a change in moving costs ($\beta$), when these are respectively fixed and variable. The results show that model outcome remains qualitatively unchanged with respect to the main analysis (see for comparison Subsection \ref{sec:movingcosts}, and Figures \ref{fig:increasing_05_fc} and \ref{fig:increasing_05_vc}): i) there is a monotonic relation between agents' level of responsiveness and the number of individual movements; ii) model outcome tends to mimic Schelling's result when the role of moving cost becomes least significant (i.e. high values of $\beta$); iii) there is a negative relation between moving costs and social welfare; iv) when agents are least subject to the constraints of moving costs, because $\beta$ exceeds a certain threshold value, simulations always reproduce the same segregation dynamics in terms of both number of iterations and moving people, and of segregation and welfare indexes: i.e., the final configuration of the grid is unaffected by any further increase of $\beta$. Only a significant change occurs in our findings, and that is when considering the case of fixed costs. Here we register a dramatic downshift in the threshold value of $\beta$ over which considered metrics become constant: i.e. from $\beta > 0.75$ in the main analysis to $\beta > 0.10$. The reason is that moving costs are fixed and close to zero, hence they have substantially no impact on the relocation choices of the agent: i.e. agents are free to choose every empty cell regardless of its distance, because their utility level remains substantially unaltered by moving costs. The same is not true when considering variable moving costs. This is because costs increases with distance, and thus they still represents a constraint for agents. As a result, the model outcome when considering low variable moving costs mimics the case considered in the main analysis, and no meaningful changes are observed.
	
\begin{figure}[H]
        \begin{center}
            \includegraphics[width=0.65\textwidth]{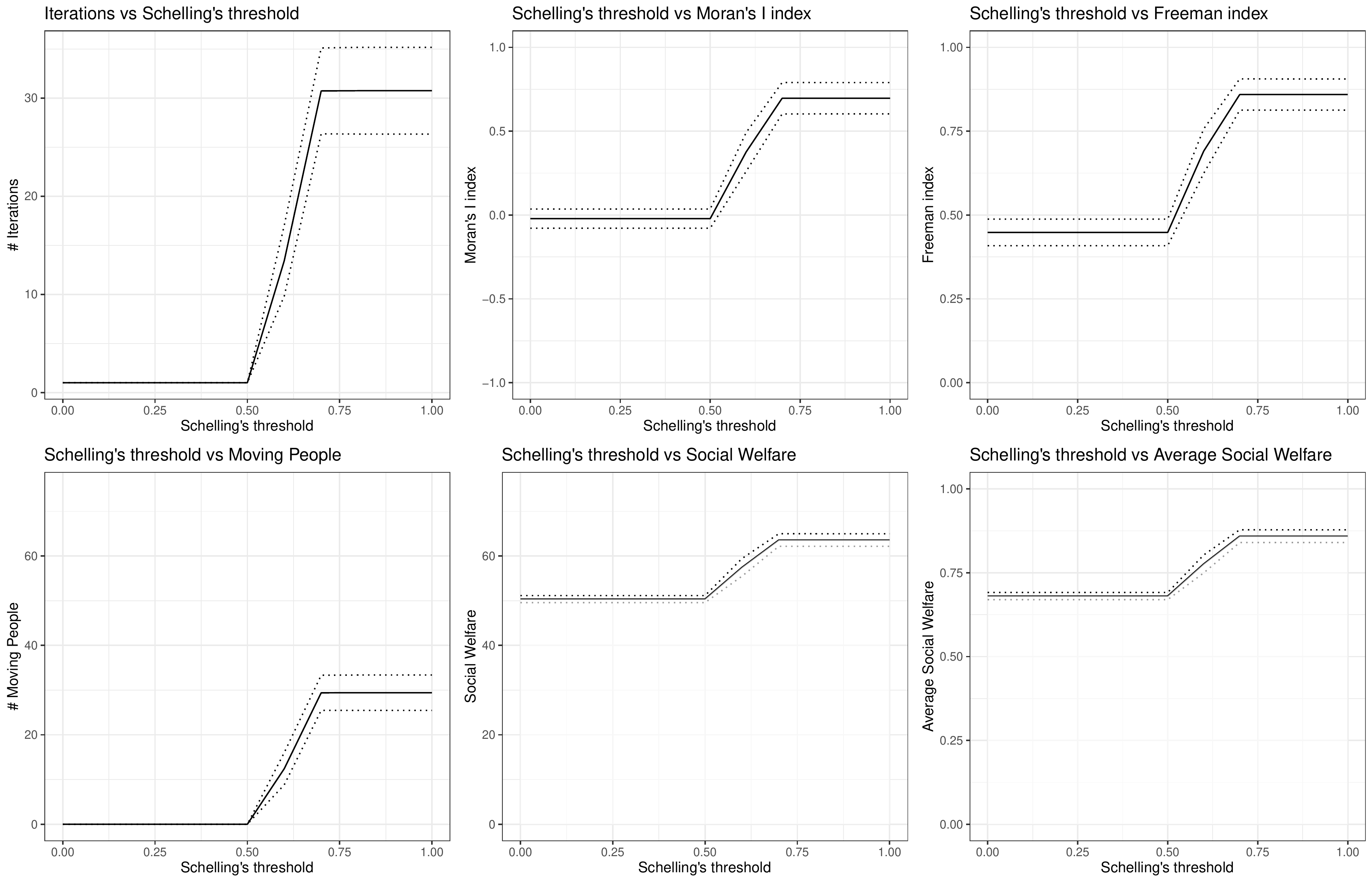}
            \caption[caption]{\scriptsize{Low fixed moving costs ($\beta = 0.5,\ \gamma = 1,\ \bar{c} = 0.01$) - No contribution of friendship to utility ($\alpha=1$).\\\hspace{\textwidth}Each iteration is repeated 100 times by permuting the initial torus. On the abscisses, we have Schelling's threshold $x$ ranging in $[0,1]$.  Bold and ticked lines indicate respectively average value and standard deviation. Upper left panel indicates the number of iterations required for the model to converge. Upper middle and right panels report respectively the Moran's I and the FSI value obtained after model convergence. Bottom left panel shows the number of people who moved at least once in a simulation. Bottom middle and right panels display respectively the average and the total social welfare of the agents at the end of the iteration process. Social welfare is obtained from the component $\beta \alpha U_i^{color}(\bar{v};x)$ of $i$-th  agent utility function (equation (\ref{UF})).}}
            \label{fig:threshold_001_fc}
        \end{center}
    \end{figure}

\begin{figure}[H]
        \begin{center}
            \includegraphics[width=0.65\textwidth]{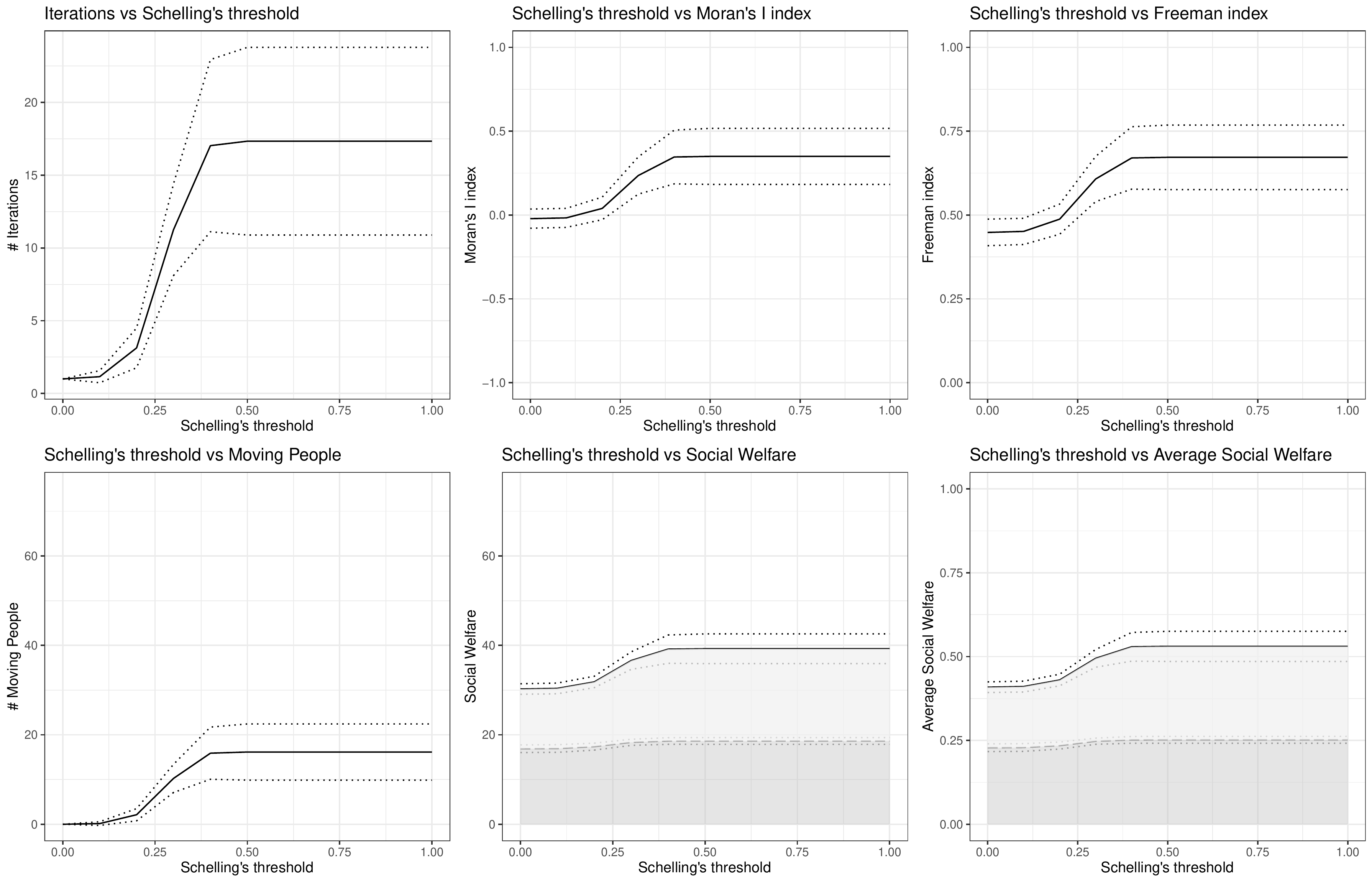}
            \caption[caption]{\scriptsize{Low fixed moving costs ($\beta = 0.5,\ \gamma = 1,\ \bar{c} = 0.01$) - Fair contribution of friendship to utility and 3 friends ($\alpha=0.5, k = 3$).\\\hspace{\textwidth}Each iteration is repeated 100 times by permuting the initial torus. On the abscisses, we have Schelling's threshold $x$ ranging in $[0,1]$.  Bold and ticked lines indicate respectively average value and standard deviation. Upper left panel indicates the number of iterations required for the model to converge. Upper middle and right panels report respectively the Moran's I and the FSI value obtained after model convergence. Bottom left panel shows the number of people who moved at least once in a simulation. Bottom middle and right panels display respectively the average and the total social welfare of the agents at the end of the iteration process. Social welfare is obtained from the component $\beta \alpha U_i^{color}(\bar{v};x)$ of $i$-th  agent utility function (equation (\ref{UF})). The color of the area indicates the two components of agents' utility after convergence is achieved: $\beta \alpha U_i^{color}(\bar{v};x)$ (lighter-colored area) and $\beta \alpha U_i^{friend}(\bar{v})$ (darker-colored area).}}
            \label{fig:threshold_friends_001_fc}
        \end{center}
    \end{figure}

\begin{figure}[H]
        \begin{center}
            \includegraphics[width=0.65\textwidth]{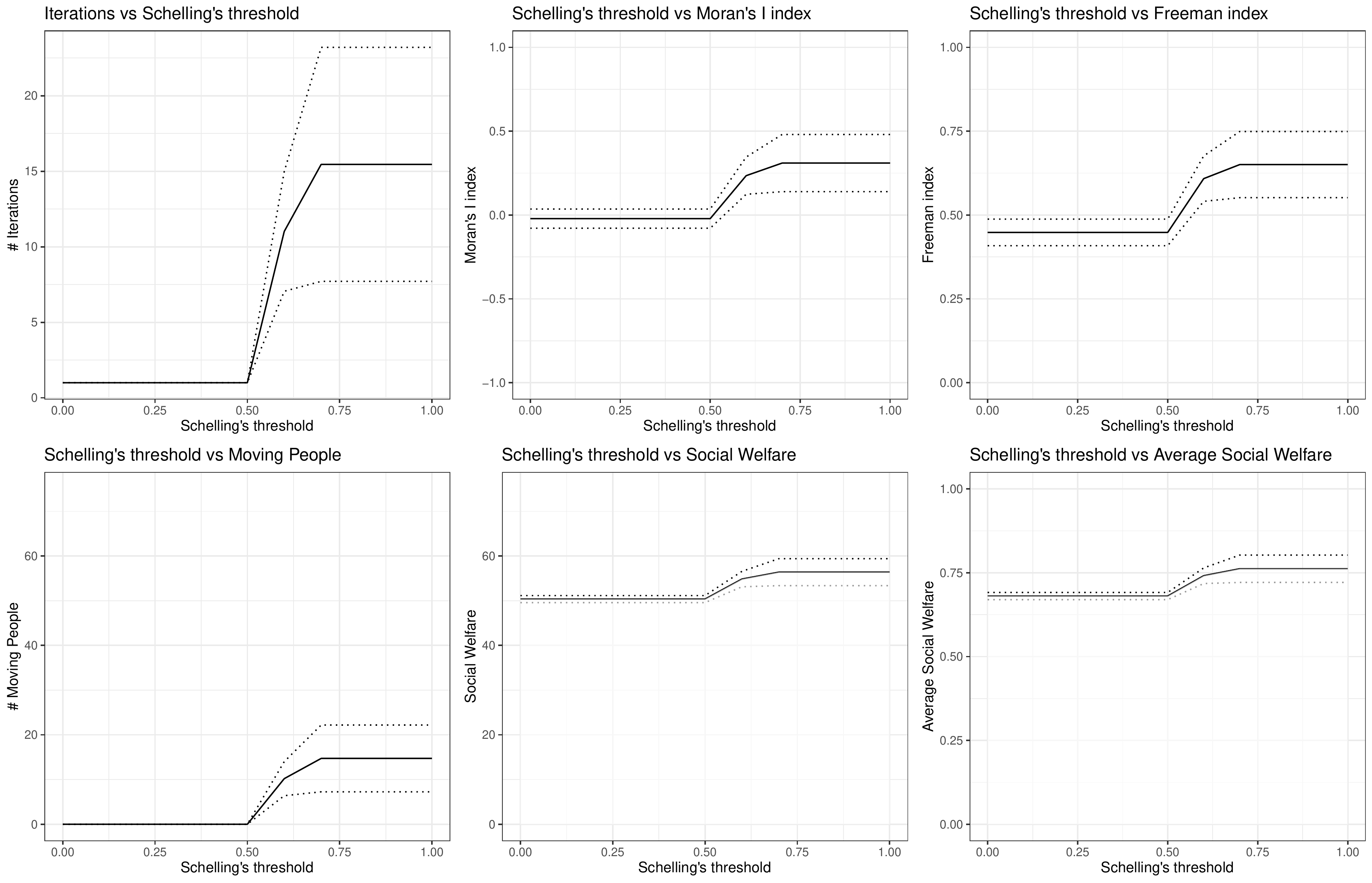}
            \caption[caption]{\scriptsize{
                    Low variable moving costs ($\beta = 0.5,\ \gamma = 0,\ \bar{c} = 0.01$) - No contribution of friendship to utility ($\alpha=1$).\\\hspace{\textwidth}Each iteration is repeated 100 times by permuting the initial torus. On the abscisses, we have Schelling's threshold $x$ ranging in $[0,1]$.  Bold and ticked lines indicate respectively average value and standard deviation. Upper left panel indicates the number of iterations required for the model to converge. Upper middle and right panels report respectively the Moran's I and the FSI value obtained after model convergence. Bottom left panel shows the number of people who moved at least once in a simulation. Bottom middle and right panels display respectively the average and the total social welfare of the agents at the end of the iteration process. Social welfare is obtained from the component $\beta \alpha U_i^{color}(\bar{v};x)$ of $i$-th  agent utility function (equation (\ref{UF})).}}            
\label{fig:threshold_001_vc}
        \end{center}
    \end{figure}

\begin{figure}[H]
        \begin{center}
            \includegraphics[width=0.65\textwidth]{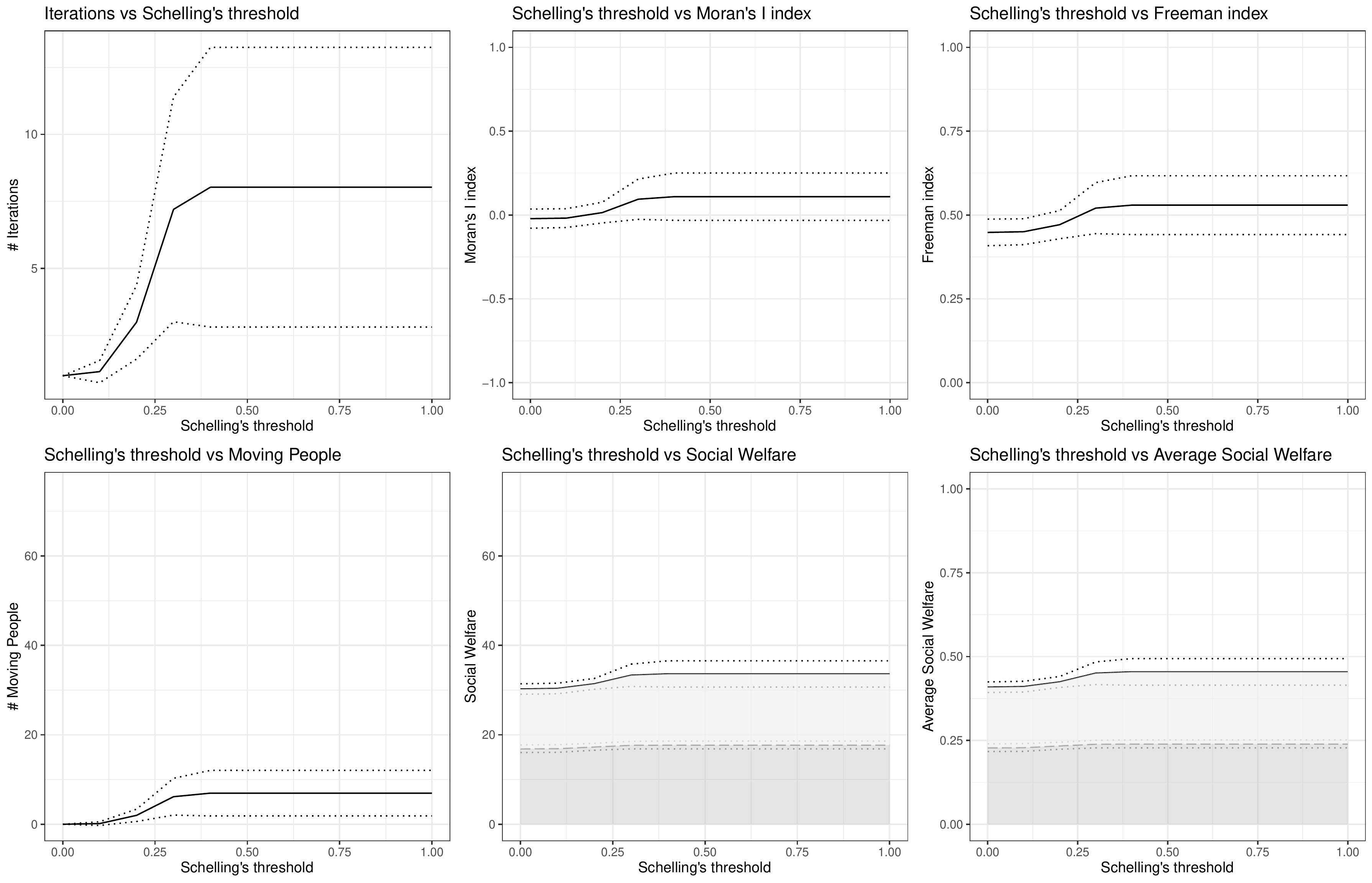}
            \caption[caption]{\scriptsize{Low variable moving costs ($\beta = 0.5,\ \gamma = 0,\ \bar{c} = 0.01$) - Fair contribution of friendship to utility and 3 friends ($\alpha=0.5, k = 3$).\\\hspace{\textwidth}Each iteration is repeated 100 times by permuting the initial torus. On the abscisses, we have Schelling's threshold $x$ ranging in $[0,1]$.  Bold and ticked lines indicate respectively average value and standard deviation. Upper left panel indicates the number of iterations required for the model to converge. Upper middle and right panels report respectively the Moran's I and the FSI value obtained after model convergence. Bottom left panel shows the number of people who moved at least once in a simulation. Bottom middle and right panels display respectively the average and the total social welfare of the agents at the end of the iteration process. Social welfare is obtained from the component $\beta \alpha U_i^{color}(\bar{v};x)$ of $i$-th  agent utility function (equation (\ref{UF})). The color of the area indicates the two components of agents' utility after convergence is achieved: $\beta \alpha U_i^{color}(\bar{v};x)$ (lighter-colored area) and $\beta \alpha U_i^{friend}(\bar{v})$ (darker-colored area).}}
            \label{fig:threshold_friends_001_vc}
        \end{center}
    \end{figure}

In Figures \ref{fig:threshold_001_fc} and \ref{fig:threshold_001_vc}, we progressively increase the responsiveness of $i$ to neighborhood composition ($x$), when moving costs are respectively fixed and variable. This complements the analysis presented in Figures \ref{fig:threshold_05_fc} and \ref{fig:threshold_05_vc} in Section \ref{sec:networked}. Results remain unaffected with respect to the main analysis. The responsiveness of the model outcome to a change in $x$ is the same in the two set of exercises. When agents do not care for the level of homophily in the composition of their neighborhood ($x < 0.5$), there is no incentive to move and convergence is achieved in one iteration as a result of the presence of moving costs. The same logic applies when we replicate the previous exercise and we increase the degree centrality from $k=0$ to $k=3$, comparing Figures \ref{fig:threshold_friends_001_fc} and \ref{fig:threshold_friends_001_vc} with respectively figure \ref{fig:threshold_friends_05fc} and \ref{fig:threshold_friends_05vc}. All in all, segregation dynamics observed in this context remains unaltered when decreasing the level of moving costs.
	
\section{Figures: High moving costs} \label{sec:appendix.figures.high.cost}

	This appendix extends our analysis and presents additional results replicating the simulations in Section (\ref{sec:results}) with higher levels of moving costs ($\bar{c} = 0.99$) with respect to the main analysis ($\bar{c} = 0.5$). Please, see footnote 10 for the details of the selected parameter set.

\begin{figure}[H]
	\begin{center}
		\includegraphics[width=0.65\textwidth]{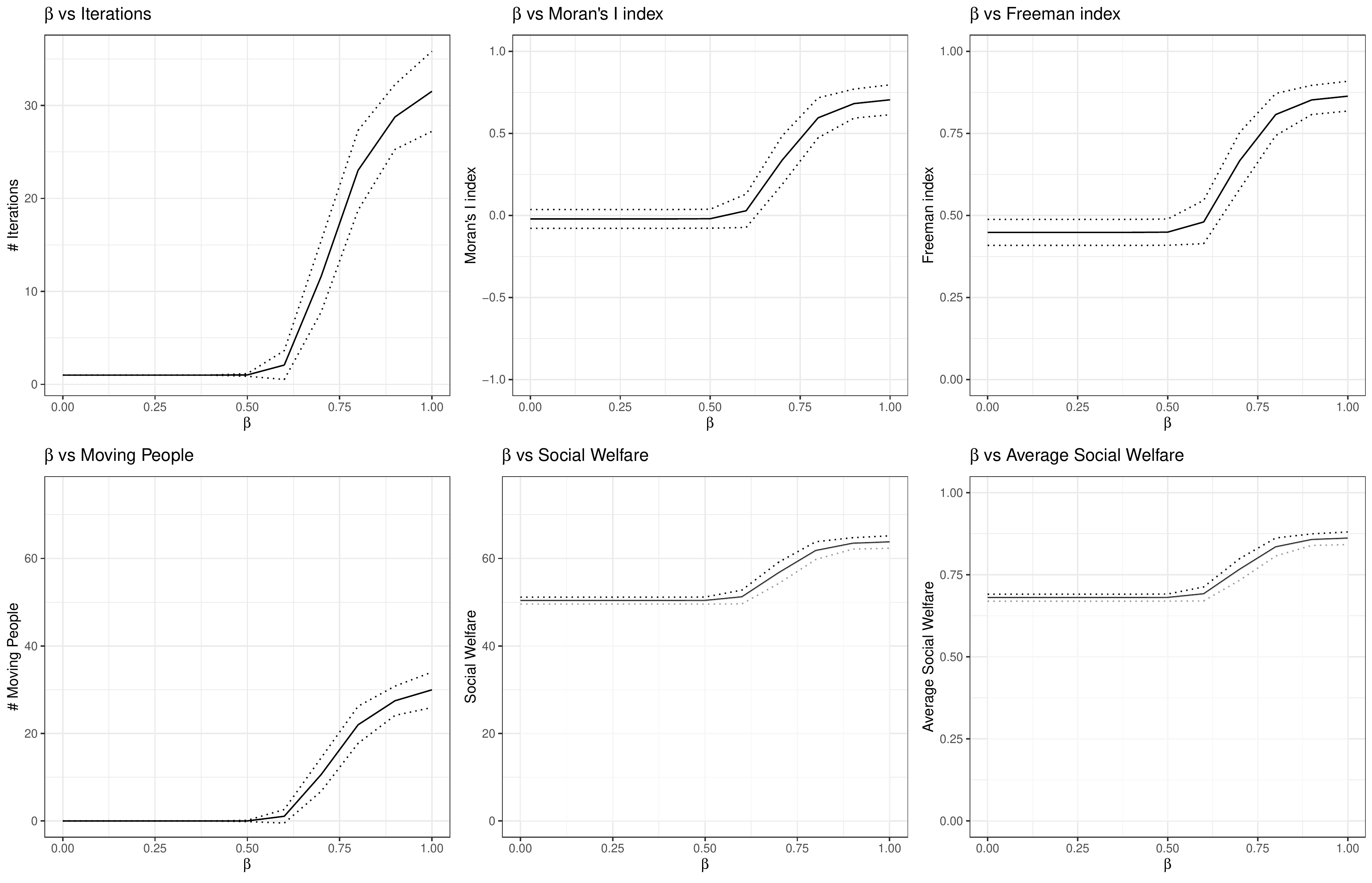}
		\caption[caption]{\scriptsize{High fixed moving costs ($\gamma = 1,\ \bar{c} = 0.99$) - Maximum value of the Schelling's threshold $x=1$ - No contribution of friendship to utility ($\alpha=1$).\\\hspace{\textwidth}Each iteration is repeated 100 times by permuting the initial torus. On the abscisses, we have parameter $\beta$ ranging in $[0,1]$.  Bold and ticked lines indicate respectively average value and standard deviation. Upper left panel indicates the number of iterations required for the model to converge. Upper middle and right panels report respectively the Moran's I and the FSI value obtained after model convergence. Bottom left panel shows the number of people who moved at least once in a simulation. Bottom middle and right panels display respectively the average and the total social welfare of the agents at the end of the iteration process. Social welfare is obtained from the component $\beta \alpha U_i^{color}(\bar{v};x)$ of $i$-th  agent utility function (equation (\ref{UF})).}}
		\label{fig:increasing_099_fc}
	\end{center}
\end{figure}

\begin{figure}[H]
	\begin{center}
		\includegraphics[width=0.65\textwidth]{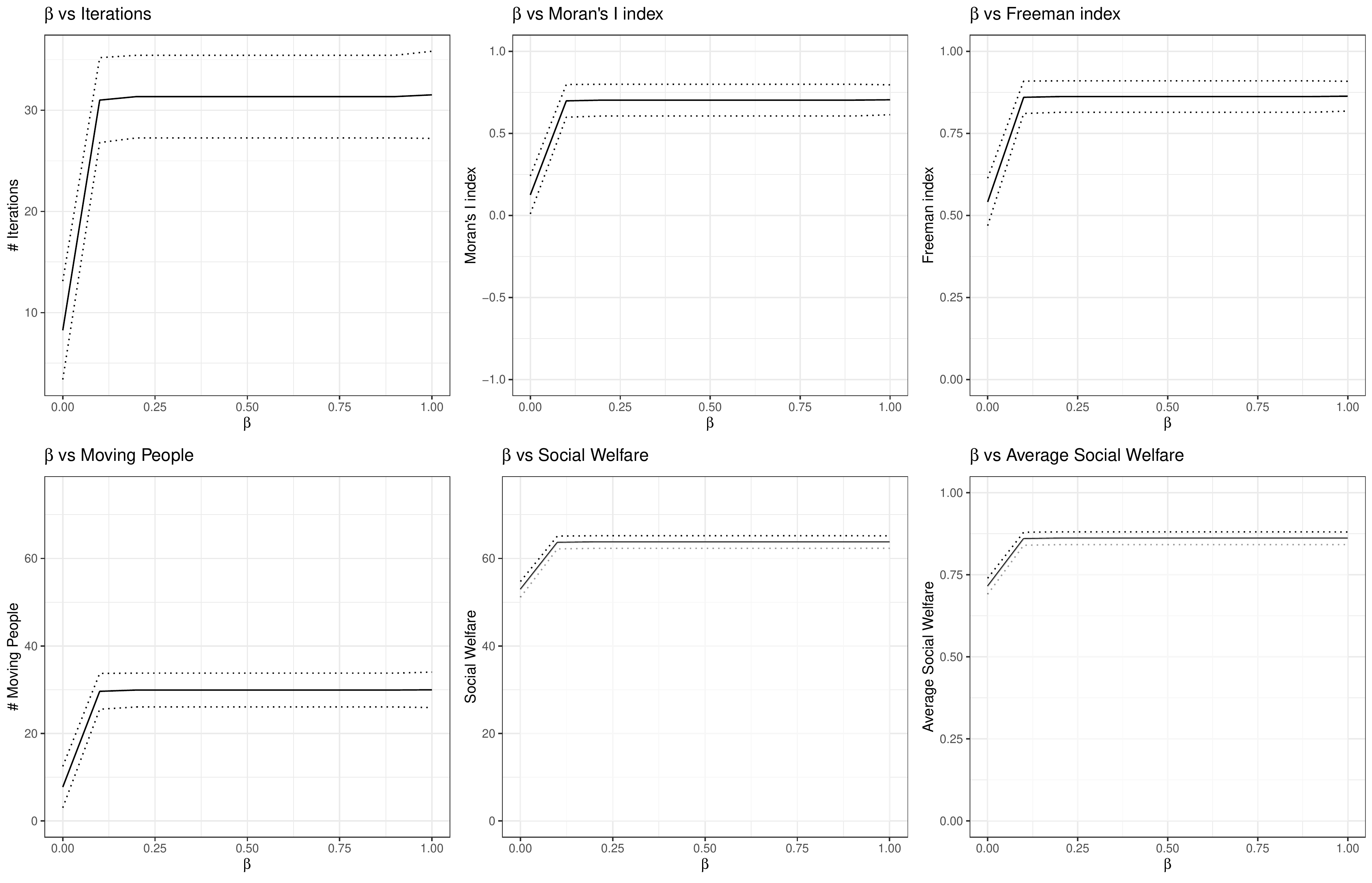}
		\caption[caption]{\scriptsize{High variable moving costs ($\gamma = 0,\ \bar{c} = 0.99$) - Maximum value of the Schelling's threshold $x=1$ - No contribution of friendship to utility ($\alpha=1$).\\\hspace{\textwidth}Each iteration is repeated 100 times by permuting the initial torus. On the abscisses, we have parameter $\beta$ ranging in $[0,1]$.  Bold and ticked lines indicate respectively average value and standard deviation. Upper left panel indicates the number of iterations required for the model to converge. Upper middle and right panels report respectively the Moran's I and the FSI value obtained after model convergence. Bottom left panel shows the number of people who moved at least once in a simulation. Bottom middle and right panels display respectively the average and the total social welfare of the agents at the end of the iteration process. Social welfare is obtained from the component $\beta \alpha U_i^{color}(\bar{v};x)$ of $i$-th  agent utility function (equation (\ref{UF})).}}
		\label{fig:increasing_099_vc}
	\end{center}
\end{figure}

	In Figures \ref{fig:increasing_099_fc} and \ref{fig:increasing_099_vc}, we progressively increase the responsiveness of $i$'s utility to a change in moving costs ($\beta$), when these are respectively fixed and variable. The results show that most model outcome remains qualitatively unchanged with respect to the main analysis when considering fixed moving costs (see for comparison Subsection \ref{sec:movingcosts} and Figure \ref{fig:increasing_05_fc}): i) there is a monotonic relation between agents' level of responsiveness and the number of individual movements; ii) model outcome tends to mimic Schelling's result when the role of moving cost becomes least significant (i.e. high values of $\beta$); and iii) there is a negative relation between moving costs and social welfare. There is only a significant difference with respect to the main analysis, and this is the level of responsiveness of the model outcome when $\beta$ is low and agents are most subject to moving costs. In these cases, there are no incentives to move: i.e. the process stops after one simulation and the initial configuration of the grid remains unaltered. When considering the case of variable moving costs instead (see for comparison Subsection \ref{sec:movingcosts} and Figure \ref{fig:increasing_05_vc}), things change radically. The choice to relocate has such a negative impact on agents' utility in this context, that relocation constraints are substantially the same regardless of the value of $\beta$. Agents move in the same way, and the same segregation dynamics are observed whenever $\beta > 0$: i.e., both number of iterations and moving people, and of segregation and welfare indexes remain constant regardless of the value of the parameter.
	
	For comments on the rest of the figures contained in this appendix, the interested reader may refer to the end of Section \ref{sec:networked}, which already contains a discussion on the significant differences between the main analysis and the case of high fixed costs.

        \begin{figure}[H]
        \begin{center}
            \includegraphics[width=0.65\textwidth]{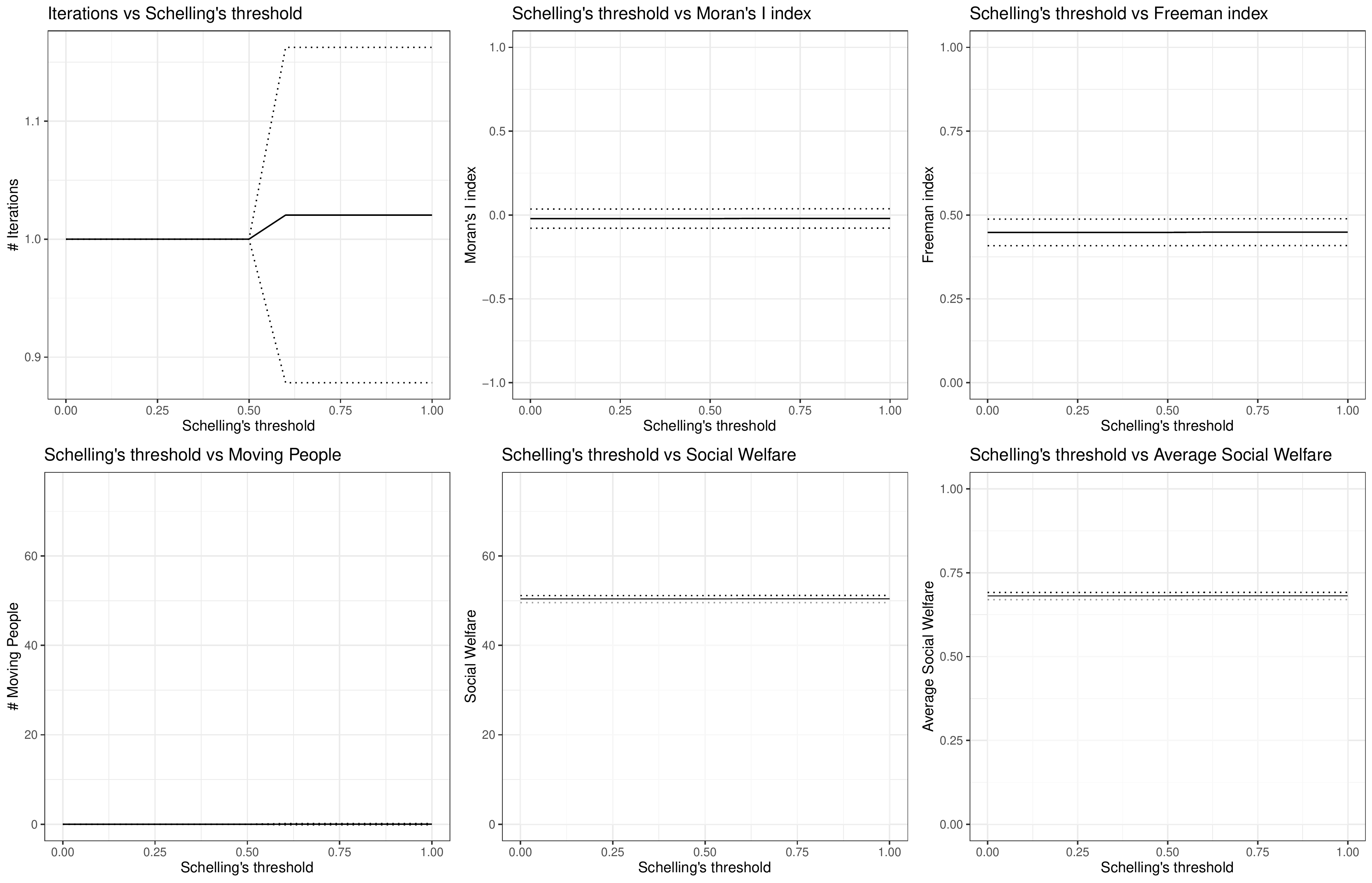}
            \caption[caption]{\scriptsize{High fixed moving costs ($\beta = 0.5,\ \gamma = 1,\ \bar{c} = 0.99$) - No contribution of friendship to utility ($\alpha=1$).\\\hspace{\textwidth}Each iteration is repeated 100 times by permuting the initial torus. On the abscisses, we have Schelling's threshold $x$ ranging in $[0,1]$.  Bold and ticked lines indicate respectively average value and standard deviation. Upper left panel indicates the number of iterations required for the model to converge. Upper middle and right panels report respectively the Moran's I and the FSI value obtained after model convergence. Bottom left panel shows the number of people who moved at least once in a simulation. Bottom middle and right panels display respectively the average and the total social welfare of the agents at the end of the iteration process. Social welfare is obtained from the component $\beta \alpha U_i^{color}(\bar{v};x)$ of $i$-th  agent utility function (equation (\ref{UF})).}}
            \label{fig:threshold_099_fc}
        \end{center}
    \end{figure}

    \begin{figure}[H]
        \begin{center}
            \includegraphics[width=0.65\textwidth]{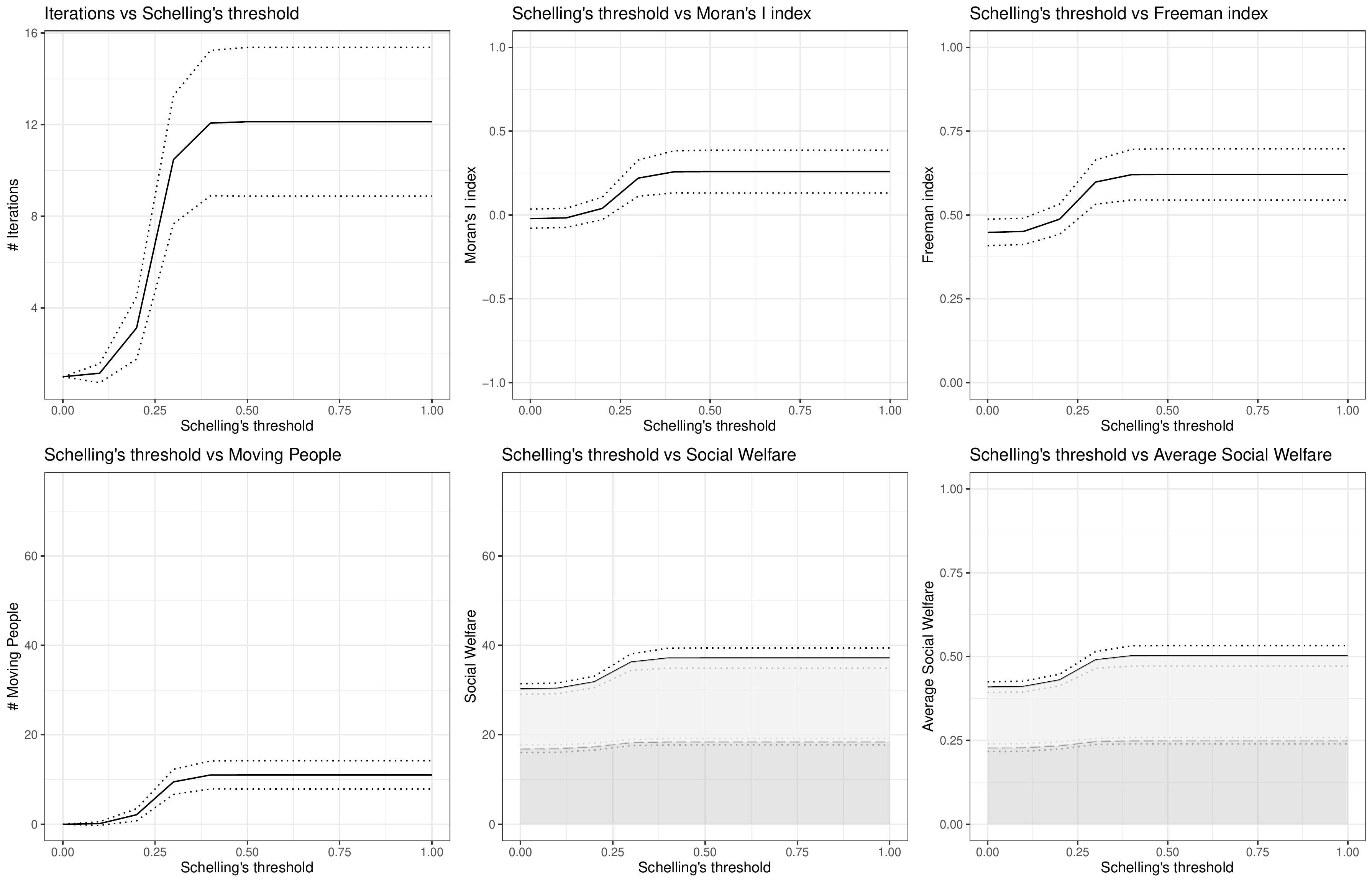}
            \caption[caption]{\scriptsize{High fixed moving costs ($\beta = 0.5,\ \gamma = 1,\ \bar{c} = 0.99$) - Fair contribution of friendship to utility and 3 friends ($\alpha=0.5, k = 3$).\\\hspace{\textwidth}Each iteration is repeated 100 times by permuting the initial torus. On the abscisses, we have Schelling's threshold $x$ ranging in $[0,1]$.  Bold and ticked lines indicate respectively average value and standard deviation. Upper left panel indicates the number of iterations required for the model to converge. Upper middle and right panels report respectively the Moran's I and the FSI value obtained after model convergence. Bottom left panel shows the number of people who moved at least once in a simulation. Bottom middle and right panels display respectively the average and the total social welfare of the agents at the end of the iteration process. Social welfare is obtained from the component $\beta \alpha U_i^{color}(\bar{v};x)$ of $i$-th  agent utility function (equation (\ref{UF})). The color of the area indicates the two components of agents' utility after convergence is achieved: $\beta \alpha U_i^{color}(\bar{v};x)$ (lighter-colored area) and $\beta \alpha U_i^{friend}(\bar{v})$ (darker-colored area).}}

            \label{fig:threshold_friends_099_fc}
        \end{center}
    \end{figure}

    \begin{figure}[H]
        \begin{center}
            \includegraphics[width=0.65\textwidth]{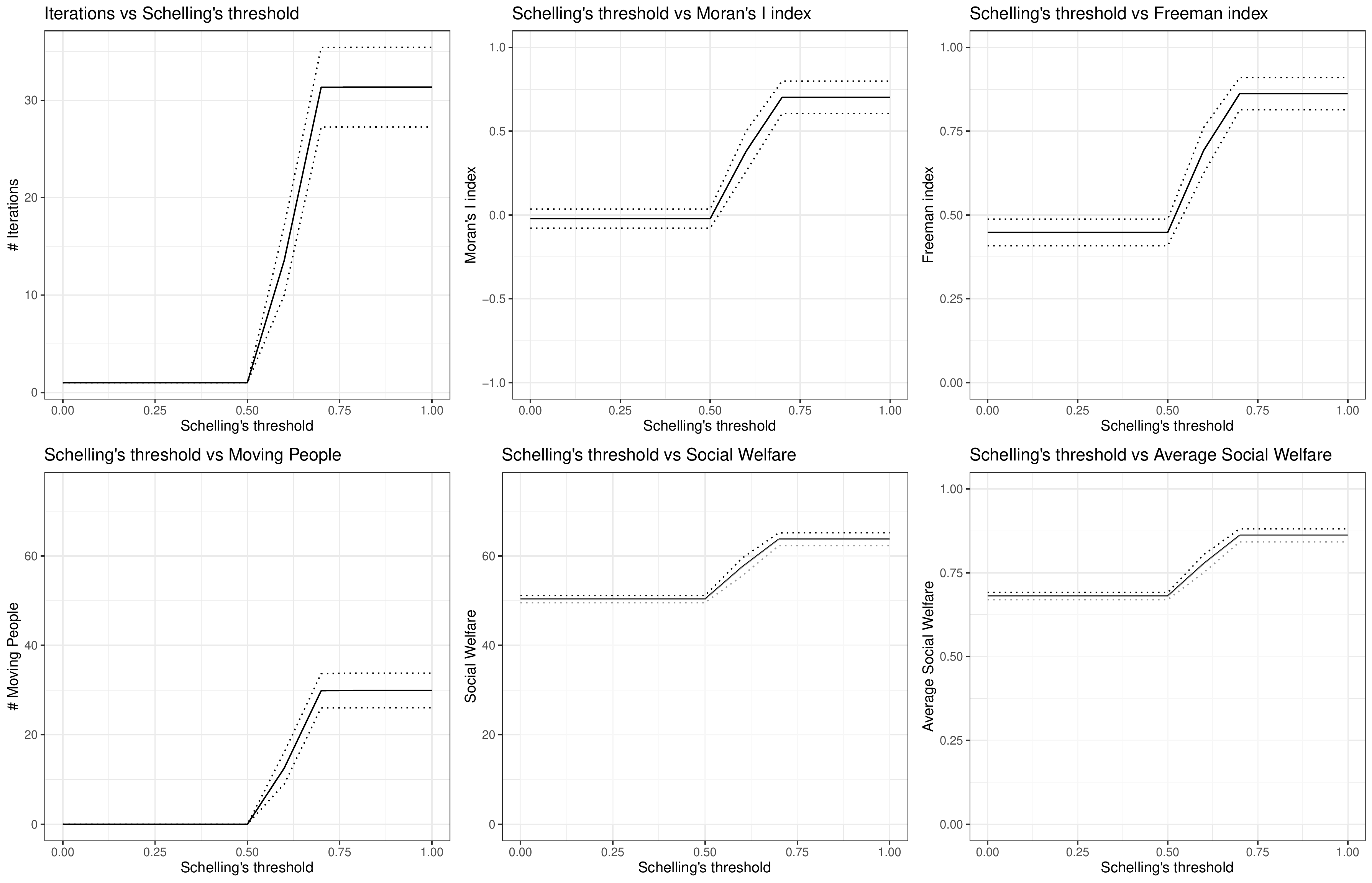}
            \caption[caption]{\scriptsize{High variable moving costs ($\beta = 0.5,\ \gamma = 0,\ \bar{c} = 0.01$) - No contribution of friendship to utility ($\alpha=1$).\\\hspace{\textwidth}Each iteration is repeated 100 times by permuting the initial torus. On the abscisses, we have Schelling's threshold $x$ ranging in $[0,1]$.  Bold and ticked lines indicate respectively average value and standard deviation. Upper left panel indicates the number of iterations required for the model to converge. Upper middle and right panels report respectively the Moran's I and the FSI value obtained after model convergence. Bottom left panel shows the number of people who moved at least once in a simulation. Bottom middle and right panels display respectively the average and the total social welfare of the agents at the end of the iteration process. Social welfare is obtained from the component $\beta \alpha U_i^{color}(\bar{v};x)$ of $i$-th  agent utility function (equation (\ref{UF})).}}
            \label{fig:threshold_099_vc}
        \end{center}
    \end{figure}

    \begin{figure}[H]
        \begin{center}
            \includegraphics[width=0.65\textwidth]{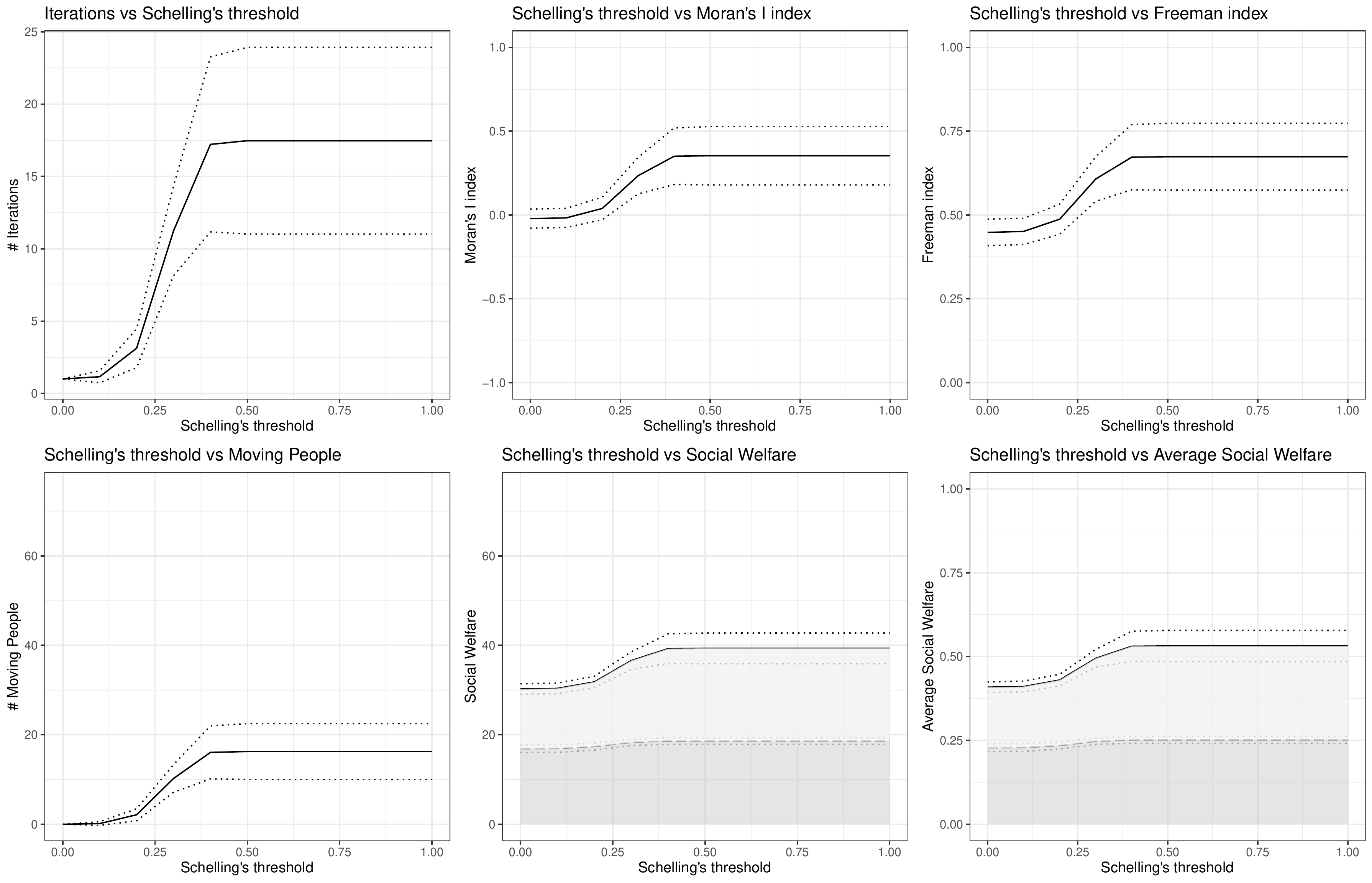}
            \caption[caption]{\scriptsize{High variable moving costs ($\beta = 0.5,\ \gamma = 0,\ \bar{c} = 0.01$) - Fair contribution of friendship to utility and 3 friends ($\alpha=0.5, k = 3$).\\\hspace{\textwidth}Each iteration is repeated 100 times by permuting the initial torus. On the abscisses, we have Schelling's threshold $x$ ranging in $[0,1]$.  Bold and ticked lines indicate respectively average value and standard deviation. Upper left panel indicates the number of iterations required for the model to converge. Upper middle and right panels report respectively the Moran's I and the FSI value obtained after model convergence. Bottom left panel shows the number of people who moved at least once in a simulation. Bottom middle and right panels display respectively the average and the total social welfare of the agents at the end of the iteration process. Social welfare is obtained from the component $\beta \alpha U_i^{color}(\bar{v};x)$ of $i$-th  agent utility function (equation (\ref{UF})). The color of the area indicates the two components of agents' utility after convergence is achieved: $\beta \alpha U_i^{color}(\bar{v};x)$ (lighter-colored area) and $\beta \alpha U_i^{friend}(\bar{v})$ (darker-colored area).}}
            \label{fig:threshold_friends_099_vc}
        \end{center}
    \end{figure}

\newpage

\bibliographystyle{ldb}
\bibliography{Segregation}

\end{document}